# Classification and Characterization of Core Grid Protocols for Global Grid Computing

Harshad B. Prajapati and Vipul K. Dabhi

*Abstract*—Grid computing has attracted many researchers over a few years, and as a result many new protocols have emerged and also evolved since its inception a decade ago. Grid protocols play major role in implementing services that facilitate coordinated resource sharing across diverse organizations. In this paper, we provide comprehensive coverage of different core Grid protocols that can be used in Global Grid Computing. We establish the classification of core Grid protocols into i) Grid network communication and Grid data transfer protocols, ii) Grid information security protocols, iii) Grid resource information protocols, iv) Grid management protocols, and v) Grid interface protocols, depending upon the kind of activities handled by these protocols. All the classified protocols are also organized into layers of the Hourglass model of Grid architecture to understand dependency among these protocols. We also present the characteristics of each protocol. For better understanding of these protocols, we also discuss applied protocols as examples from either Globus toolkit or other popular Grid middleware projects. We believe that our classification and characterization of Grid protocols will enable better understanding of core Grid protocols and will motivate further research in the area of Global Grid Computing.

*Index Terms*—Grid Protocols, Classification of Grid protocols, Characterization of Grid protocols, Grid information security protocols, Grid resource information protocols, Grid interface protocols, Grid management protocols, Global Grid Computing.

## I. INTRODUCTION

Towards the realization of Global Grid Computing, many researchers have been attracted to broad research areas such as architectures, protocols, services, APIs, and toolkits of Grid in a last decade. Grid applications exploit services provided by Grid architecture to enable various scientific applications such as high performance computing, high-throughput computing, distributed super-computing, data-intensive computing, etc. Different services in Grid computing are defined in terms of the protocols that are used to interact with them and the expected behaviors from the services. Considerable research is going on protocols used in Grid computing and certain protocols are also standardized by Global Grid Forum [1]. However, proprietary Grid protocols are used for domain/application specific Grid architectures. Interoperability among these protocols is a headache while integrating different proprietary Grid architectures.

The three point checklist definition of Grid in [2], "*Grid is a system that coordinates resources that are not subject to central control using standard, open, general-purpose protocols and interfaces to deliver nontrivial qualities of service*", stresses on standard protocols and interfaces to enable global deployment of Grid as it is today for the Internet. Many international Grid projects (e.g., [3], [4], [5]) and national Grid projects (e.g., [6], [7], [8]), started in a last decade, have demonstrated the power and capability of Grid computing and have stimulated Grid computing at global scale to solve different scientific problems. Moreover, the distributed.net [9] and the SETI@home [10] popularized the Grid computing at global scale. The availability of Grid middleware (e.g., Globus) has opened doors for other forms of Grid computing such as cloud computing, utility computing, autonomic computing, volunteer computing, etc. In this paper, by Global Grid Computing, we mean Grid be able to exploit resources provided by any organizational, national, or international Grids that exist today or will exist in future. Consequently, to enable this global connectivity for resource sharing at global scale, Foster in [2] suggests requirement of common, standardized, InterGrid Grid protocols for any entity to become part of global Grid.

In simplest words, a protocol can be defined as a communication procedure with the specific rules governing the syntax, semantics, sequence of statements, and synchronization of communication between communicating entities. A protocol in computing includes set of rules for (i) formatting the messages that are exchanged, (ii) message exchange sequence/pattern, and (iii) timings of exchanged messages. Certain protocols in computing can also specify the behavior as a result of particular message exchange. Behavior specification enables different implementations of behavior possible thus not restricting the creativity. Grid protocols play major role in implementing services that facilitate Global Grid Computing. We expect the standardization of all Grid protocols that are used in implementing various activities and services for global Grid deployment.

H. B. Prajapati is with the Information Technology Department of Faculty of Technology, Dharmsinh Desai University, Nadiad 387001, Gujarat, INDIA. (e-mail: harshad.b.prajapati@gmail.com).
V. K. Dabhi is with the Information Technology Department of Faculty of Technology, Dharmsinh Desai University, Nadiad 387001, Gujarat, INDIA.

Considerable research is still to be carried out for well engineering and standardizing the Grid protocols. Our goal in the paper is to provide good understanding and comprehensive coverage of major core Grid protocols that can be used for Global Grid Computing by classify and characterize approach.

Our objectives in this paper are (i) Classify all core Grid protocols that can be used in Global Grid Computing into different categories depending upon the kind of activities handled by the protocols. (ii) Organize all classified protocols into Hourglass model of Grid architecture to understand protocol dependency among them. (iii) Characterize all core Grid protocols to understand specific characteristics and behaviors of the protocols. We believe that our classification and characterization of core Grid protocols will enable better understanding of core Grid protocols and will motivate further research in the area.

The paper is structured as follows. Section II discusses related work and motivation for the work. Section III discusses basic concepts, terms, and related terminology. Section IV provides the classification of core Grid protocols that can be used for Global Grid Computing and also shows placement of classified core Grid protocols in Hourglass model based Grid architecture. Section V characterizes and discusses protocols related to Grid network communication and Grid data transfer. Section VI characterizes and discusses protocols related to Grid information security. Section VII characterizes and discusses protocols related to Grid resource information. Section VIII characterizes and discusses protocols related to Grid management activities. Section IX characterizes and discusses protocols related to providing Web services standards based interfaces. Finally, Section X provides the conclusions. We have tried to make this paper approachable to the reader unfamiliar with the Grid, however, the reader interested in getting exhaustive details need to read relevant referenced work.

## II. Related Work and Motivation for the Work

The work in [11] on "The core protocol set for the Global Grid" addresses only communication protocols. This work addresses protocols at transport layer (TCP, UDP), protocols at network layer (IP, ICMP, IGMP, and various routing protocols), and protocols at application layer (FTP, SMTP, Telnet). The work provides good discussion on protocols related to network communication; however, we believe that Grid inherently requires certain basic and core services such as data transfer, job management, scheduling, security, etc without which Grid computing is not possible. The protocols related to these basic activities are not addressed in the work. The work in the "Layered communications architecture for the Global Grid [12]" addresses layered reference model for Global Grid without touching various activities required in Grid computing and related protocols. Our work in this paper will comprehensively address Grid protocols and will enable better understanding of various core Grid protocols.

In this paragraph, we discuss motivation for our work. Our preliminary reading on Grid protocols indicated that number of different protocols exists in Grid to handle various Grid activities and interactions. Moreover, our initial investigations in current Globus toolkit–GT4 revealed that it is Service Oriented Architecture (SOA) based. And programming of Grid applications on GT4 is done using Web services based interfaces. However, any service functionality of Grid inherently requires use of protocols (messages, timing, formats, etc.). Moreover, with the use of SOA based Grid, programmers get benefit of using uniform interfaces to access various Grid services. Thus, only message formats and their transports have changed, while other constituent of earlier Grid protocols still play a major role internally in making Grid API based on Web services. As the classification method has always remained a fundamental method in science to understand activities and characteristics of a large set of various entities, we use that method to discuss a number of different core Grid protocols that can be used for Global Grid Computing. As many production systems still use earlier non-SOA based Grid middleware (e.g., Globus Toolkit–GT2) and fundamental concepts of earlier traditional Grid have influenced current Grid architecture, we feel it appropriate to cover protocols that are found in SOA based Grid and earlier traditional Grid.

## III. Fundamental Concepts, Terms, and Terminology

In this section, we discuss basic concepts, terms, and related terminology, which will help the readers to understand the work in this paper. Using this fundamental concepts, terms, and terminology, we also establish the context of the content in this paper.

- *Grid*: It is a system that coordinates distributed resources using standard, open, general-purpose protocols and interfaces to deliver non-trivial qualities of service [13].
- *Global Grid*: A Global Grid is a worldwide computing environment, offering a diverse range of heterogeneous services managed and owned by self-interested parties [14]. Global Grid can offer various kinds of services such as compute service, data services, application services, interaction service, knowledge service and it can support distributed supercomputing, high throughput computing, content-sharing, data-intensive computing, on-demand or real-time computing, collaborative computing, and so on [14].
- *OGF-Open Grid Forum (GGF-Global Grid Forum):* It is an international forum or community of users, developers, and leading vendors that produces the specifications (for services, protocols, components, etc.) and best practices to enable international Grid efforts [1].
- *Virtual Organization (VO):* It is a set of individuals and/or institutions defined by the rules of coordinated resource sharing [15]. Diversity in local access policies of contributed resources places major challenges in realizing such VOs as uniformly accessible.

- *Protocol*: It is a complete and unambiguous set of rules governing message formats, their syntax and semantics, message timing, message error handling, and message synchronization of communication between communicating entities [16].
- *Grid Protocol*: It is a protocol that mediates between user and resource during access to any resource in Grid environment.
- *Service*: A service is a network-enabled entity with a well-defined interface that provides some capability to its clients by exchanging messages [17]. In Grid, a service is defined in terms of the protocol that is used to interact with it and the behavior expected in response to various protocol message exchanges (i.e., service=protocol+behavior) [15].
- *Resource*: A resource is an entity that may be shared and exploited in a networked environment [18]. Resources such as computational resources, storage resources, network resources, sensors, and instruments can be shared using Grid for scientific, engineering, high performance computing applications.
- *Globus toolkit*: It is community-based, open-architecture based, open source set of services and software components that support Grid and Grid applications [19]. These software components include Grid Security Infrastructure – GSI protocols (as information security mechanism), Monitoring and Discovery Service – MDS (as information service), Grid Resource Allocation Manager – GRAM (for remote job management), and GridFTP (as data transfer protocol).

While mentioning about Globus toolkit in this paper, wherever version specific details are provided, it is indicated with version number (e.g., GT-2 or GT-4), and wherever version number is not provided, the reader can assume that the discussed concepts are equally applicable to GT-2 and GT-4.

## IV. CLASSIFICATION OF CORE GRID PROTOCOLS AND THEIR PLACEMENT IN HOURGLASS MODEL BASED GRID ARCHITECTURE

The goal of Global Grid is supporting diverse heterogeneous resources at global scale and providing worldwide computing with offering various kinds of services. To support heterogeneous resources and to offer heterogeneous services require generic interfaces for resources and services, respectively. The Grid information service, Grid monitoring service, Grid scheduling and brokering service, Grid security services, and Grid data transfer service are basic services that must be supported by Global Grid [14]. In this paper, we discuss important protocols related to these basic/generic services.

Various protocols are designed and few existing protocols are also adapted in Global Grid Computing to make overall Grid infrastructure modular, extensible and scalable. We classify different Grid protocols into five main categories: (i) Grid network communication and Grid data transfer protocols, (ii) Grid information security protocols, (iii) Grid resource information protocols, (iv) Grid management protocols, and (v) Grid interface protocols, depending upon kind of activity handled by these protocols. The full classification of these protocols is shown in Fig. 1.

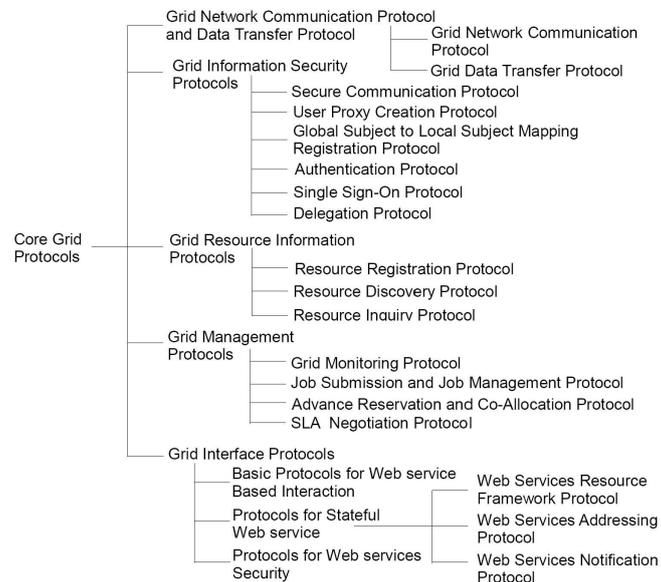

Figure 1 Classification of core Grid protocols for Global Grid Computing. The classification is based on kind of activities handled by the protocols.

We, first, briefly describe the Hourglass model based Grid architecture presented in [13]. We, then, show placement of all discussed protocols in appropriate layers of this Grid architecture to understand dependency among these protocols. The Hourglass model contains small set of core abstractions and protocols (e.g., resource and connectivity protocols) at neck utilized by high level behaviors (i.e., collective layer services) and implementable on a wide range of resource types (i.e., different fabrics). This open and extensible Grid architecture is shown in Fig. 2. This Grid architecture identifies required fundamental system components to implement a Grid and also specify purpose, function, role, responsibility of these components. Each layer has well-defined protocols and interfaces that provide access to different defined services.

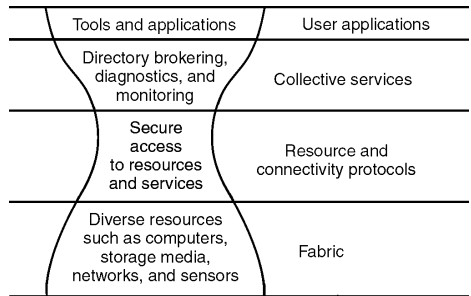

Figure 2 Hourglass model based Grid architecture. (Source, [13])

This Grid architecture has five layers named as (i) fabric, (ii) connectivity protocols, (iii) resource protocols, (iv) collective services, and (v) user applications. The components in the Grid Fabric layer implement the local, resource specific operations that occur on specific resources. The connectivity layer defines core communication and authentication protocols required for Grid-specific network interactions and transactions. The resource layer builds on the connectivity layer, and this layer defines the protocols for secure negotiation, initiation, monitoring, control, accounting, and payment of sharing operations on *individual resources*. The resource and connectivity layer protocols form the neck of this Hourglass model. Thus, the protocols in these two layers should be limited in number and generic enough to support building diverse functionality at above layer. The collective layer contains protocols and services not associated with any one specific resource but instead associated with *collections of resources*. This layer can include services such as directory services, co-allocation service, scheduling service, brokering service, and *monitoring and diagnosis* services. The collective layer protocols range from general purpose to highly application or domain specific protocols. The final, topmost layer is an application layer which contains applications that operate within Grid environment. The constructed applications can use services defined at any layer, including application layer also where application/domain specific services can be placed.

Important core Grid protocols are organized into layers of Hourglass model based Grid architecture to understand protocol dependency among them. Table I shows placement of core Grid protocols into the layers (collective, resource, and connectivity) of Hourglass model based Grid architecture. The table also shows applied protocols as examples from either Globus toolkit or other popular middlewares. The dependency of the layers and hence constituent protocols is like this: the collective layer protocols depend on resource layer protocols, and resource layer protocols depend on connectivity layer protocols.

TABLE I PLACEMENT OF CORE GRID PROTOCOLS IN HOURGLASS MODEL BASED GRID ARCHITECTURE

| Layer | Protocol | Example/Applied protocol |
|---|---|---|
| Application | Any | - |
| Collective | Resource discovery protocol | GRid Information Protocol (GRIP) of Globus toolkit |
| | Co-allocation protocol | RealityGrid's Co-allocation protocol |
| | SLA negotiation protocol | Service Negotiation and Acquisition Protocol (SNAP) |
| | Grid monitoring protocol | GRid Information Protocol (GRIP) of Globus toolkit |
| Resource | Resource registration protocol | GRid Registration Protocol (GRRP) of Globus toolkit |
| | Resource inquiry protocol | GRid Information Protocol (GRIP) of Globus toolkit |
| | Grid monitoring protocol | GRid Information Protocol (GRIP) of Globus toolkit |
| | Grid Data transfer protocol | GridFTP protocol of Globus toolkit |
| | Advance Reservation protocol | RealityGrid's advance reservation protocol |
| | Job submission and job management protocol | Grid Resource Allocation and Management (GRAM) protocol of Globus toolkit |
| Connectivity | Secure communication protocol | Transport Layer Security/Secure Socket Layer (TLS/SSL) protocol of Internet |
| | User proxy creation protocol | Proxy certificate based user proxy creation protocol of Globus toolkit |
| | Global subject to local subject mapping registration protocol | Global subject to local subject mapping registration protocol of Globus toolkit |
| | Authentication protocol | X.509 certificate and proxy certificate based Authentication protocol of Globus toolkit |
| | Single Sign-On (SSO) protocol | Proxy certificate based Single Sign-On (SSO) protocol of Globus toolkit |
| | Delegation protocol | Proxy certificate based Delegation protocol of Globus toolkit |
| | Grid Communication protocols | Transmission Control Protocol (TCP), User Datagram Protocol (UDP) of Internet |
| Fabric | Any | - |

For realizing SOA based global Grid, Web services standards provide required standard based, open, interoperable protocols and specifications. The protocol dependency among Web services protocols is shown in Fig. 3. The Web service stack, as shown in Fig. focuses on only important core Web services protocols for SOA based Grid; however there exist other Web services protocols and specifications.

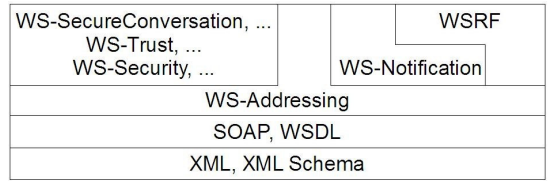

Figure 3 Protocols dependency in Web service stack. Three protocols WS-Notification, Web Service Resource Framework (WSRF), and Web services security can be used independently.

The future's Grid architecture will be based on open, standard based, interoperable interfaces. The future's SOA based Grid architecture exploiting Web services protocols is shown in Fig. 4. The Fig. also shows interaction of applications with various layers of traditional Grid architecture and Future's SOA based Grid architecture. The future's Grid architecture will provide standard interfaces to various layers of Hourglass model based Grid architecture using OGSA interfaces. The OGSA interfaces for various core Grid services are being standardized by OGSA WG of Global Grid Forum.

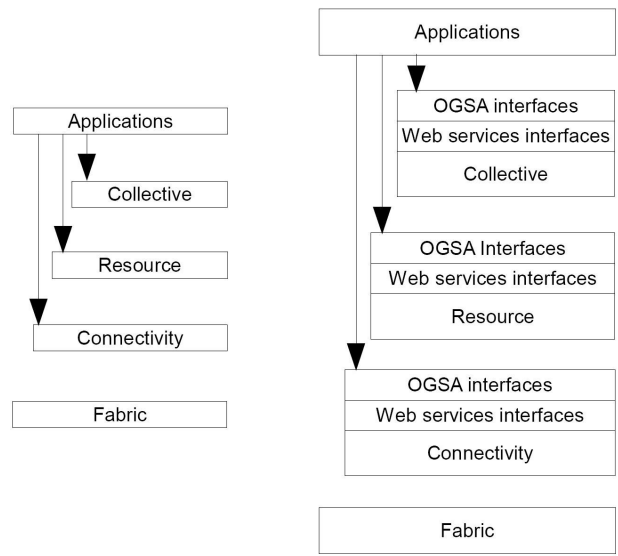

Figure 4 Comparison of applications' interaction in traditional Grid architecture and future's Service Oriented Architecture based Grid architecture. OGSA interfaces are being standardized by Open Grid Services Architecture Working Group (OGSA-WG) of Global Grid Forum. For creating Web services interfaces, the protocols of Web service stack, as shown in Figure 3, are used.

## V.  GRID NETWORK COMMUNICATION AND GRID DATA TRANSFER PROTOCOLS

The characteristics of Grid network communication protocol and Grid data transfer protocol are summarized in Table II. The table shows at-the-glance look of these two protocols, however, comprehensive discussion on the protocols is provided in the following sub-sections.

TABLE II CORE GRID NETWORK COMMUNICATION AND DATA TRANSFER PROTOCOLS, THEIR USAGE, AND THEIR CHARACTERISTICS

| Protocol | Example | Protocol Usage | Characteristics of Protocol |
|---|---|---|---|
| Grid Network Communication Protocol | TCP of Internet | To provide reliable end to end delivery of application data | • Byte stream oriented transfer of data.<br>• Ordered delivery of packets.<br>• Reliable communication establishment and termination.<br>• Congestion control mechanism. |
| | UDP of Internet | To provide unreliable end to end delivery of application data | • Unordered and unreliable delivery of data packets.<br>• Uni-cast, broadcast and multicast communication. |
| Grid Data Transfer Protocol | GridFTP of GT | To provide fast, secure, efficient, reliable transfer of large data among Grid nodes | • Secure data transfer using Grid Security Infrastructure (GSI) and Kerberos based authentication.<br>• Third-party control of data transfer.<br>• Parallel, stripped, and partial data transfer.<br>• Automatic negotiation of TCP buffer/window sizes.<br>• Reliable data transfer by restarting failed transfer. |

### A. Grid Network Communication Protocol

The ultimate goal of Grid computing is allowing sharing of resources across Grid nodes in networked environment. In Grid environment, transport of voice, video, data, control and commands makes realization of higher level services and applications possible. At the core of any form of communication among Grid participants, the network communication protocol transports the messages and data over network. Other higher level services and protocols are built using network communication protocols. The network communication protocol needs to consider different requirements of users. Data reliability, error detection and recovery, flow control, uni-cast communication, broadcast communication, multi-cast communication, wireless communication, etc. would be requirements of higher level Grid services and protocols. TCP/IP protocol stack is commonly used for network communication.

*Example Protocol: Transmission Control Protocol (TCP)*

TCP [20] is a transport layer protocol of a TCP/IP protocol stack, which is widely used and matured protocol suite for Internet. TCP is reliable and byte stream oriented end-to-end protocol. It supports reliable and error free communication between end hosts by handling different issues such as reliable connection establishment, reliable connection termination, end-to-end flow control and network congestion control transparently from end users. It runs on top of network layer of the TCP/IP protocol stack. TCP does not assume anything about underlying communication systems and can be used with systems varying from Ethernet, dial-up telephone lines, to large-scale networks like the Internet.

*Example Protocol: User Datagram Protocol (UDP)*

UDP [21] is a transport layer connection-less protocol that provides datagram based (unlike stream based), unreliable and unordered message delivery. It is a light weight version of TCP and it does not handle issues as mentioned above in TCP. UDP uses a simple transmission model without guaranteeing reliability and ordering of data packets, and handling congestion control in the network. UDP supports multicast and broadcast message delivery also. Time-sensitive applications such as audio, video transfer or broadcast are often implemented using UDP due to its distinguished characteristics.

Interested readers are directed to [22] for getting details on other available network communication protocols.

### B. Grid Data Transfer Protocol

The Grid environment needs a fast, secure, efficient, and reliable transport mechanism to transfer data among Grid nodes due to its distributed nature. The Grid requires protocol for transferring job and associated data on executor node and gathering result of the completed job and for transferring large amount of data between storage systems. Moreover, the data transfer protocol should be able to handle errors and faults (e.g., transient network failures, server outages, etc) in networked environment. Moreover, the data transfer among Grid nodes needs to consider security of data and users.

*Example Protocol: GridFTP of Globus Toolkit*

GridFTP [23] is a common data transfer protocol that provides secure and efficient data movement in Grid environment. This protocol extends the standard FTP protocol [24] to provide support for parallel, stripped, and partial data transfers [25]. We briefly present here the characteristics of GridFTP as discussed in [23]. However, interested readers are directed to [25] [26] for getting more details on GridFTP.

(i) Secure data transfer: GridFTP supports both Grid Security Infrastructure (GSI) and Kerberos based authentication mechanisms. GridFTP provides this capability by implementing the Generic Security Services(GSS)API authentication mechanisms defined by RFC 2228, "FTP Security Extensions" in [27]. (ii) Third-party control of data transfer: GridFTP provides third-party control of transfers between storage servers. (iii) Parallel data transfer: GridFTP improves aggregate bandwidth by using multiple TCP streams. (iv) Stripped data transfer: GridFTP uses multiple TCP streams to transfer data that is partitioned across multiple servers. (v) Partial data transfer: Many Grid applications require the transfer of partial files. GridFTP supports new FTP commands, as presented in [25], to support transfers of regions of a file. (vi) Automatic negotiation of TCP buffer/window sizes: GridFTP extends the standard FTP command set and data channel protocol to support both manual setting and automatic negotiation of TCP buffer sizes both for large files and large sets of small files. (vii) Support for reliable data transfer: Reliable data transfer is important for many data intensive scientific applications. The GridFTP protocol exploits features specified by FTP standard to provide reliability and fault recovery by restarting the failed data transfer.

## VI. GRID INFORMATION SECURITY PROTOCOLS

Security problem in network communication is very complex to handle. Thus, it becomes important to exploit existing standard based protocols for security solutions in Grid environment, wherever possible. Many security standards developed for Internet protocol suite become applicable for security solution in Grid. However, due to distinct characteristics of Grid, it requires solutions to Grid specific security requirements. Apart from basic security requirements such as authentication, authorization, integrity and confidentiality, other important security requirements in Grid are non-repudiation, single sign on (SSO), delegation, credential life span and renewal, integration with local security solutions, policy exchange, secure logging, assurance, manageability, firewall traversal, and user based trust relationships [28] [29].

Due to existence of diverse organizational security policies, a Grid requires a global unique identity of Grid participants and associated credentials for security operations. The Grid handles local heterogeneity of resources by using X.509 certificate [30].

We briefly discuss here important concepts of credential and certificate. A Credential is a piece of information (eg. passwords or certificates) that is used to prove the identity of a subject. Grid uses X.509 certificate [30] as a global unique identity of subject. The X.509 certificate is a tamper-proof digital identity that is cryptographically signed by a well-know Certificate Authority (CA). X.509 certificate contains a unique distinguished name (DN) and other related information about the individual user or host that is being certified. The DN is formed using hierarchical naming scheme, as used in X.500 [31], containing organization name (O), organization unit name (OU) and common name (CN), etc. In Grid environment, the certificate can be issued to user (user certificate) as well as to Grid node or service (called host certificate or service certificate).

The characteristics of Grid information security protocols are summarized in Table III. The table shows at-the-glance look of these protocols, however, comprehensive discussion on the protocols is provided in the following sub-sections.

TABLE III CORE GRID INFORMATION SECURITY PROTOCOLS, THEIR USAGE, AND THEIR CHARACTERISTICS

| Protocol | Example | Protocol Usage | Characteristics of Protocol |
|---|---|---|---|
| Secure Communication Protocol | TLS/SSL of Internet | To provide confidentiality, data integrity, and authentication (server authentication, client authentication, and mutual authentication) for secure network communication. | • X.509 certificate based authentication.<br>• Negotiation for protocol version, cipher settings (algorithm, key length), and authentication.<br>• Symmetric cryptography for encryption-decryption of application data.<br>• Asymmetric cryptography for secure exchange of session key. |
| Authentication Protocol | X.509 certificate and proxy certificate based authentication protocol of GT | To verify identity of Grid entity. | • X.509 certificate as a global credential of Grid entity.<br>• Proxy certificate as a global credential of dynamic Grid entity.<br>• Secure communication through TLS/SSL. |
| User Proxy Creation Protocol | Proxy certificate based user proxy creation protocol of GT | To create a proxy process with proxy credential that can work on behalf of (in absence of) user. | • Secure creation of proxy process<br>• Uses proxy certificate as a temporary credential for proxy process. |
| Global Subject to Local Subject Mapping Registration Protocol | Proxy certificate based global subject to local subject mapping registration protocol of GT | To create and manage mapping between global subject and local subject | • Stores mapping information in grid-map file.<br>• Creation of global subject to local subject mapping without involvement of administrator.<br>• Secure and reliable mapping creation process. |
| Single Sign-On (SSO) Protocol | Proxy certificate based single sign-on protocol of GT | To authenticate user once and allow performing activities without re-authentication. | • Uses proxy process with proxy certificate as a credential.<br>• Uses private key associated with user's long term credential only once. Uses proxy certificate and associated private key for frequent re-authentication.<br>• Secures private key associated with user's long term credential. |
| Delegation Protocol | Proxy certificate based delegation protocol of GT | To securely delegate subset of rights to dynamic grid entity. | • Uses TLS/SSL protocol for secure communication.<br>• Uses proxy certificate to include delegation rights in form of X.509 extension called Proxy Certificate Information (PCI).<br>• Allows usage of any arbitrary policy expression language (e.g., Keynote, XACML, XrML, etc).<br>• Can limit path length of delegation to Grid entities. |

### A. Secure Communication Protocol

As Grid involves networked nodes, and/or institutions, and/or organizations, it requires mechanism to exchange information securely. A secure communication requires (i) confidentiality (data privacy) − data should be accessible only to the party the data is intended for, i.e., protecting the sensitive data from prying eyes while it is in transit, (ii) message integrity (tamper proofing of data) − ensuring that the unauthorized changes made to message content or data can be detected at the other end of communication, and (iii) user authentication − making sure a user is who he claims to be.

*Example Protocol: Secure Socket Layer/Transport Layer Security Protocol (SSL/TLS)*

TLS [32] is a dominant security protocol in the Internet. It can secure any type of traffic above stream oriented transport protocol (i.e., TCP). It provides confidentiality, data integrity, and authentication (i.e., server authentication, client authentication, and mutual authentication). TLS supports X.509 certificates [30] and uses public key cryptography (PKI) [33] for authentication and secure key distribution. TLS uses symmetric cryptography for secure exchange of application data. The symmetric cryptography in TLS requires secure exchange of symmetric cryptography key, and it is achieved using TLS handshaking. Using TLS handshaking, the server and client can authenticate each other and they can agree on encryption algorithm, cryptographic keys, and their length, etc.

TLS has three sub-protocols (i) Handshake Protocol, (ii) Change Cipher Spec Protocol, and (iii) Alert Protocol. The Handshake Protocol is responsible for negotiating about the secure session related information. The Change Cipher Spec Protocol is used to signal transitions in ciphering strategies. The Alert Protocol is used to convey warning or fetal messages. The Alert Protocol also specifies what action should be taken when some alert message arrives.

In use of TLS protocol, before any data is securely exchanged between two networked entities (i.e., client and server), TLS handshaking occurs between these two entities. In the process, both entities agree on a TLS protocol version; select cryptographic algorithm; optionally authenticate each other; and use public-key encryption techniques to generate shared secrets (i.e., random numbers and pre-master key). Both entities generate a secret master key from shared secrets using same algorithm. Once both communication entities share a common secret session key after negotiation, the key is used as a symmetric key for encryption and decryption of data which is to be exchanged between two entities.

### B. User Proxy Creation Protocol [28]

A user proxy is an entity that acts on user's behalf. It is defined by Foster in [28] as *"A user proxy is a session manager process given permission to act on behalf of a user for a limited period of time."* This protocol creates a temporary credential for the user proxy process. This temporary credential includes validity interval and other restrictions imposed by the user (e.g., host names, target sites). The steps of user proxy creation protocol, shown in Fig. 5, are as follows.

Step 1: The user logons to the Grid node on which user proxy process is to be created using whatever form of local authentication is placed on that node. Step 2: A new key-pair of public-private keys is created on Grid node. Step 3: A new certificate (credential), called proxy certificate, is created using public key (created in step 2) and is signed using private key of the user. Step 4: A new process is created on the Grid node. Step 5: The newly created process is assigned proxy certificate and associated private key. The created process is called user proxy process. Step 6: The user proxy process can use this proxy certificate (credential) to prove its identity to the requester (authenticator) using process as discussed in authentication protocol.

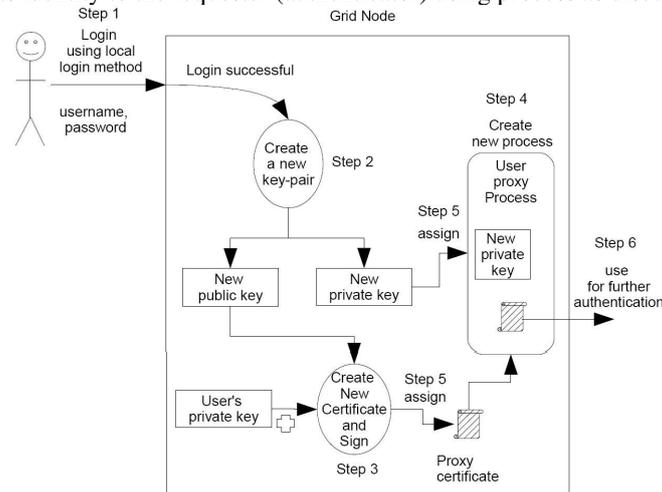

Figure 5 Certificate based user proxy creation protocol

### Example Protocol: Proxy Certificate Creation Protocol in Globus Toolkit

Proxy certificate [34] is used as a credential for user proxy process in Grid. The subject name of proxy certificate is similar to its owner's. However, while generating subject name of proxy certificate, /CN=proxy is appended to the subject name of its owner. A proxy certificate is created by generating a new public-private key pair and by signing the public key by using user's own credential instead of involving a CA. This mechanism allows new credentials and identities, required for dynamic Grid entities (e.g., job or process), to be created quickly without the involvement of a traditional administrator. The Grid Security Infrastructure (GSI) − a component of Globus toolkit, has an implicit policy that any two entities bearing proxy certificates issued by the same user will inherently trust each other. This policy allows users to create trust domains dynamically by issuing proxy certificates to any services that they wish to have collaboration.

### C. Global Subject to Local Subject Mapping Registration Protocol

Grid applications are characterized by dynamic resource requirements, use of resources from multiple administrative domains and complex communication structures. Moreover, at the architecture point of view, the sharing mechanisms in VO cannot change local security policies and organization's resource access rules. Moreover, the mechanisms for access must allow individual institutions or organizations to maintain control over their own resources. To satisfy these Grid security requirements [28], correct mapping between a global subject and a corresponding local subject is required. The resource proxy process is responsible for maintaining and managing the mappings for different users. A resource proxy is defined by Foster in [28] as *"an agent used to translate between interdomain security operations and local intradomain mechanism."* The mappings are maintained in mapping table. The mappings can be created by a local system administrator. However, to avoid any error while doing mapping (if done manually by a local system administrator), the mapping protocol is used to create the mapping at resource proxy. The basic idea in this protocol is that the user securely informs to the resource proxy process, through authentication, that it holds both global credential and local credential. As the resource proxy process can accept both global and local credentials, the user can securely assert about the mapping to resource proxy process by following both global authentication and local authentication methods respectively.

*Example Protocol: Global Subject to Local Subject Mapping Registration Protocol of Globus Toolkit*

The Globus toolkit uses Grid Resource Allocation Manager (GRAM) as resource proxy and proxy certificate as credential for user proxy process. The Globus toolkit stores mapping between global subject and local subject in grid-map-file, a file on each Grid node that user requires access to. The steps in Global Subject to Local Subject Mapping Registration Protocol of Globus Toolkit, shown in Fig. 6, are as follows.

Step 1: The User proxy process authenticates with the resource proxy process. Step 2: And then, the user proxy process creates a signed request $\text{MAP}_{\text{User Proxy}}\{\text{global subject} \rightarrow \text{resource subject}\}$ using its private key. Step 3: The user proxy process issues this mapping request to the resource proxy. Step 4: The user logs on to the resource (e.g., Grid node) using the resource's local authentication method. Step 5: Once login is successful, the user starts a map registration process. Step 6: This map registration process issues a request $\text{MAP}_{\text{Registration Process}}\{\text{global subject} \rightarrow \text{resource subject}\}$ to the resource proxy. Step 7: The resource proxy process waits for $\text{MAP}_{\text{User Proxy}}$ and $\text{MAP}_{\text{Registration Process}}$ requests with matching mapping arguments. The resource proxy process ensures (not shown in the Fig. for simplicity) that the map registration process belongs to the resource subject specified in the map request. Step 8: If the resource proxy process finds a match, it sets up a mapping in mapping table and sends acknowledgement to the map registration process and the user proxy process. Step 9: If resource proxy process does not receive a match within a certain time period (Map-timeout), it removes the pending request (not shown in the Fig.) and sends an acknowledgement to the corresponding map registration process and the user proxy process. If the user proxy process and map registration process do not receive any acknowledgement, they consider that their requests have failed.

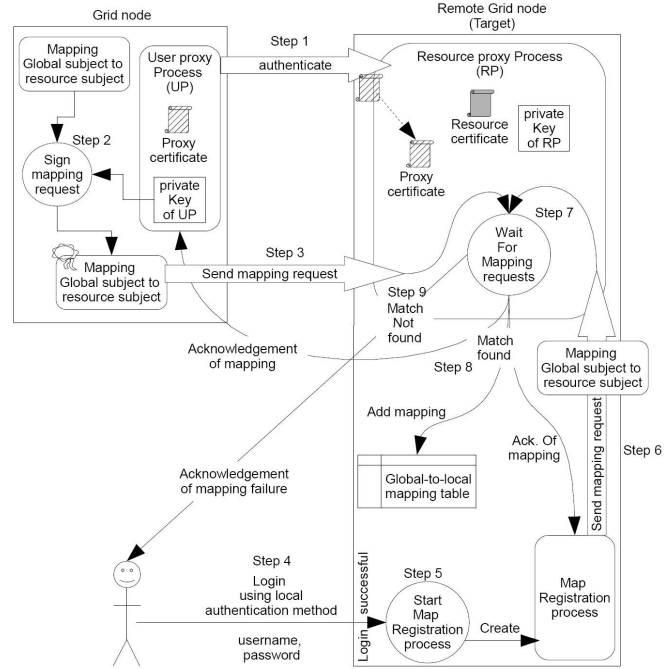

Figure 6 Global subject to local subject mapping registration protocol

### D. Authentication Protocol

Authentication is the process by which an identity of a subject is verified by requester (authenticator), typically through the use of credential [28]. Before a user can access a Grid or VO, it has to obtain its global or VO level identity (e.g., certificate) signed by CA of particular Grid or VO. The process involves creation of global identity in form of user certificate. User certificate is used as a credential while authenticating to Grid. The interesting property of private-public key pair of Public Key Infrastructure (PKI) [33] − whatever you encrypt using private key becomes available by decrypting the encrypted content using public key, and vice versa, helps in authentication process.

*Example Protocol: Certificate based Authentication Protocol of Globus Toolkit*

During the certificate based authentication process, the communication between user and authenticator is secured through TLS/SSL. The steps of certificate based authentication protocol [35], shown in Fig. 7, are discussed here.

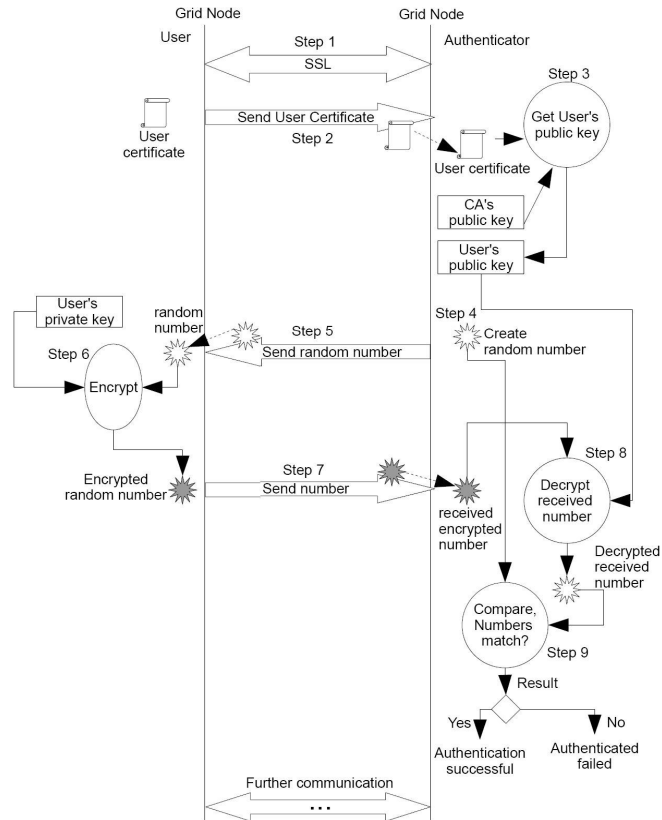

Figure 7 Certificate based authentication protocol.

Step 1: The user on one Grid node establishes a secure channel with authenticator on another Grid node. Step 2: The user sends its certificate to authenticator node. Step 3: On receiving the user's certificate, the authenticator node retrieves the public key of the user (subject) from its (i.e., the sent) certificate. Step 4: The authenticator node generates a random number. Step 5: The authenticator sends the generated random number to the user. Step 6: The user encrypts the received random number using its private key. Step 7: The user sends the encrypted number to the authenticator. Step 8: Now, the authenticator decrypts the sent encrypted random number using public key of the user, which becomes available in step 3. Step 9: The authenticator node now compares two random numbers: one, which was generated by itself in step 4 and second, which it received from the user in step 8. If these two numbers match, then the user is who he claims to be, and thus the user becomes authenticated.

*Example Protocol: Proxy Certificate based Authentication Protocol of Globus Toolkit*

When users utilize proxy credentials (e.g., proxy certificates) for authentication, the authentication protocol slightly differs. In addition to above protocol steps, the authenticator first needs to verify the signature of proxy certificate. In the process, a chain of trust is established from the CA to the proxy through the owner of the proxies. The steps of verifying signature of proxy certificate are as follows.

Step 1: The authenticator node receives proxy certificate as well as proxy's owner certificate from user. Step 2: The authenticator verifies signature of proxy certificate using public key of owner's certificate. Step 3: The authenticator then verifies signature of owner's certificate using CA's public key.

### E. Single Sign-On (SSO) Protocol

Consider a scenario in which a user starts a Grid application that needs to distribute its jobs to other remote Grid nodes and those Grid nodes in turn also need to distribute their child or sub-jobs to other Grid nodes under user's security policy. In such situation, if the user has to repeatedly provide passphrase (i.e., password placed on file containing associated private key) in order to get authenticated for different activities initiated by an application on user's behalf on different or same Grid nodes, it places burden on users and also there is a possibility of private key be compromised during repeated access. A Single Sign-On mechanism solves this problem by creating a temporary proxy credential for user and using this proxy credential rather than long term credential for repeated authentication. Single Sign-On (SSO) is an ability to get authenticated (sign-on) just once, rather than once per resource or administered domain accessed [13]. Through SSO protocol [36], user authenticates itself once, and then can perform multiple actions without re-authentication.

*Example Protocol: Proxy Certificate based Single Sign-On Protocol of Globus Toolkit*

In Certificate based PKI, the private key associated with long term credential is kept protected (using passphrase) by its owner. Moreover, the authentication using this long term credential requires owner's intervention, which does not work if owner is not available at the time of authentication. The SSO protocol avoids this restriction.

During SSO process, a user authenticates once in order to create proxy process and associated certificate. Once proxy process is created, it can be used repeatedly to authenticate for some period of time, thus enabling SSO. As discussed earlier, with proxy certificate a new public-private key pair is generated and user's long term private key is never used during any number of times re-authentication is performed using proxy certificate. The user proxy creation protocol, as discussed earlier, is used to create proxy credential and hence proxy certificate. The proxy certificate and its associated private key are stored in files which are protected only by local file system permissions to allow for easy and frequent access by the user [36]. This proxy certificate and associated private key are used by its holder to authenticate to other entities.

### F. Delegation Protocol

Delegation [36] is the act of transferring rights and privileges to another party [37]. During execution of certain Grid applications, a Grid user needs to delegate some of its rights to applications for a brief amount of time and on relatively short notice; so that the applications can continue their work without user's manual involvement. The delegation of rights is also useful in the cases when entities (e.g., services or jobs) are created dynamically. Thus, the delegation allows remote processes, jobs, services or resources to act on user's behalf.

*Example Protocol: Proxy Certificate based Delegation Protocol of Globus Toolkit*

The proxy certificate can contain X.509 extension, called Proxy Certificate Information (PCI), using which the issuer of proxy certificate can express its desire to delegate rights to the holder of proxy certificate and also to limit further creation of proxy certificate by proxy certificate holder [36]. The delegation rights can be expressed using policy expression language such as Keynote, XACML, XrML, etc. Proxy certificate allows the issuer to use any delegation policy expression it chooses with the restriction that the trusting (i.e., who gets delegated rights) party understands that policy expression [36].

PCI extension includes two fields (i) a policy method identifier (i.e., using Object Identifier – OID) and (ii) a policy itself (i.e., policy expression in selected policy method) to allow use of any arbitrary policy language to express policy expression. The proxy certificate RFC [34] mandates two policy methods (i) Proxying (i.e., delegate all rights) and (ii) Independent (i.e., delegate no rights) that must be supported by all implementations of proxy certificates. When any of these two policies is used, the policy field will be empty, as the intended delegation policy is explicit in the type. The PCI extension can also contain a field conveying the maximum path lengths (i.e., proxy creates another proxy, and so on) over which proxy certificates can be created. A value of zero indicates only initial issuer can issue the proxy certificate, i.e., proxy of first proxy is not possible. If the field is absent, it indicates path length can be unlimited.

The steps of a proxy certificate based delegation protocol, shown in Fig. 8, are as follows [36].

Step 1: The delegator on one host connects to the delegatee on remote host using SSL protocol. Step 2: Both, delegator and delegatee perform mutual authentication, delegator using proxy certificate and delegatee using its service certificate. Step 3: The delegator expresses its desire to delegate the rights using application specific method. Step 4: The delegatee generates a new public-private key pair. Step 5: By including new public key, created in above step, the delegatee creates a signed certificate request using its service's private key. Step 6: The delegatee sends that request to the delegator over secure channel for verification. Step 7: The delegator decrypts the signed request using public key of the target service. The public key of target service can easily be derived from its certificate, which becomes available in step 2. Step 8: The delegator creates a new proxy certificate using public key retrieved in Step 7. While creating a new proxy certificate, the delegator fills appropriate PCI fields (i.e., delegation rights and proxy certificate path length). Step 9: The delegator uses its private key associated with its proxy certificate to sign the new proxy certificate. Step 10: The newly created proxy certificate is sent back over the secured channel to the delegatee. Step 11: The delegatee places this received proxy certificate and associated private key in its local file system. Now, the delegatee has proxy certificate which works as delegation credential while authenticating to other required services.

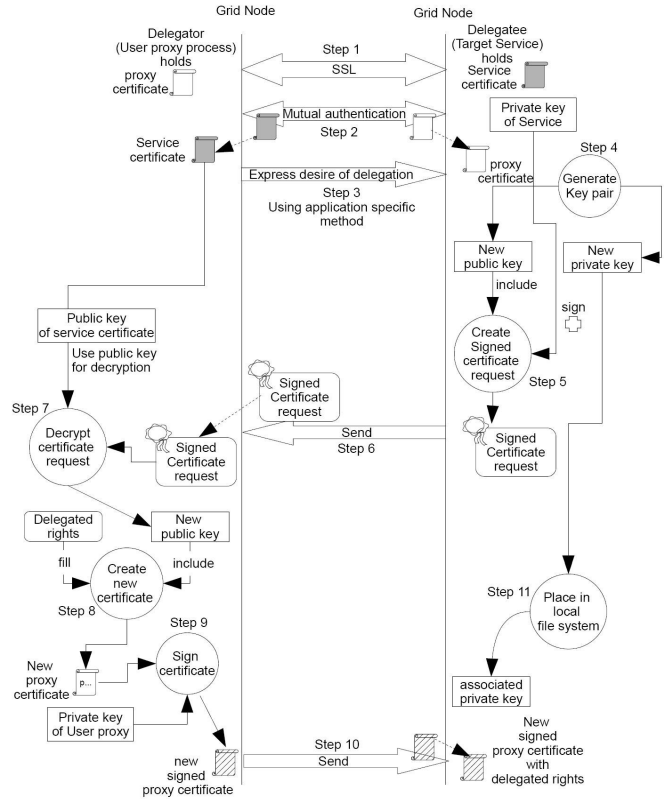

Figure 8 Proxy certificate based delegation protocol

## VII. GRID RESOURCE INFORMATION PROTOCOLS

The availability of correct information on state of Grid plays a major role while exploiting Grid resources for Grid applications. Thus, the information management is very crucial to aid resource usage to Grid applications. For this reason, three protocols, resource discovery, resource inquiry, and resource registration, are provided by the information service [38] component in Grid. The Grid resource information protocols can be combined with other Grid protocols to construct additional higher level services and capabilities such as brokering, monitoring, fault detection, and troubleshooting.

Before we discuss on information protocols, we would like to briefly mention about the information service component from [38]. The information service should be as distributed and decentralized as possible so that it becomes resilient to failure of any constituent component(s). For this reason, the Grid information service comprises two fundamental entities: highly distributed information providers and specialized aggregate directory. The Interactions among these two entities: aggregate directories and information providers and users or schedulers are shown in Fig. 9.

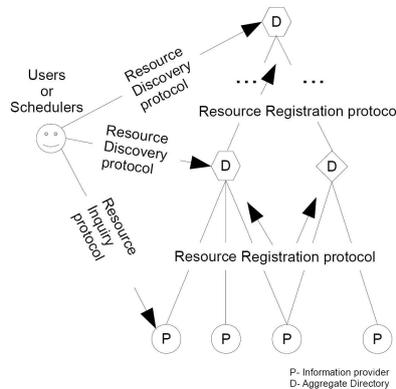

Figure 9 Interactions among users, aggregate directories, and information providers through Grid resource information protocols

- *Information Provider (IP) [38]*: It is an entity that manages information about resources shared in Grid environment. It provides access to information about individual entities (resources or services). Information Provider maintains information such as availability, capability, characteristics, load status, resource access policies, and cost of using resources. Information providers form a common VO neutral infrastructure providing access to detailed and dynamic information about Grid entities.

- *Aggregate Directory [38]*: It facilitates resource discovery and monitoring for VO by implementing both generic and specialized views and search methods for a collection of resources. It provides often specialized, VO-specific views of federated resources, services, and so on.

The characteristics of core Grid resource information protocols are summarized in Table IV. The table shows at-the-glance look of these protocols, however, comprehensive discussion on the protocols is provided in the following sub-sections.

TABLE IV CORE GRID RESOURCE INFORMATION PROTOCOLS, THEIR USAGE, AND THEIR CHARACTERISTICS

| Protocol | Example | Protocol Usage | Characteristics of Protocol |
|---|---|---|---|
| Resource Registration Protocol | GRid Registration Protocol (GRRP) of MDS-2 of GT-2. | To register information about Grid entities (resources, services, etc.) in index service. | • Soft state based registration.<br>• Uses LDAP for forming notification messages. |
| Resource Discovery Protocol | The GRid Information Protocol (GRIP) of MDS-2 of GT-2. | To search resources from index service with coarse grained matching criteria. | • Implemented as LDAP.<br>• LDAP defines a data model for message structure and format; query language for search operation; and wire protocol.<br>• Supports secure interaction using Grid Security Infrastructure (GSI). |
| Resource Inquiry Protocol | The GRid Information Protocol (GRIP) of MDS-2 of GT-2. | To inquire resource characteristics by interacting with Information Provider | • Implemented as LDAP.<br>• LDAP defines a data model for message structure and format; query language for query operation; and wire protocol.<br>• Does not require storage of resource information, i.e., information can be generated dynamically. |

### A. Resource Registration Protocol

Resource provider in the Grid is an entity that provides its resources to be usable by others. But to enable sharing, the resource provider has to make its resources or services visible to others. Using resource registration protocol, the Information Provider, which possesses information on resources, makes resources or services visible to other higher level services (e.g., directory services). For the liveliness of information and the reduction of information update overhead, the soft state communication mechanisms [39] are employed by the protocol. In soft state based registration protocol, the resource state at directory is eventually discarded unless it is periodically refreshed by resource provider.

*Example Protocol: The GRid Registration Protocol (GRRP) of MDS-2 of GT-2.*

GRid Registration Protocol (GRRP) is used by an Information Provider to notify an aggregate directory of its availability for indexing, or by an aggregate directory to invite an information provider to join a Virtual Organization (VO) [15]. GRRP is also used by a directory to notify other higher level directory about its existence, thus forming the hierarchical directory structure. GRRP is implemented as soft-state protocol [39] and it may be implemented on a variety of reliable or unreliable transport protocols. A GRRP message contains the name of the service that is being described, the type of notification message, and timestamps (for soft state mechanism) that determine the time interval over which the notification should be considered valid. GRRP in MDS-2 was implemented as Lightweight Directory Access Protocol (LDAP) [40].

### B. Resource Discovery Protocol

In Grid environment, higher level services (users or schedulers) use discovery protocol to obtain information about the Grid entities (e.g., resources or services) known to a directory service. The resource discovery is the process of querying the distributed state of the Grid to identify those resources whose characteristics and state match those desired by the resource consumer [18]. The discovery protocol provides view of many different entities in the Grid. Thus, it works at a Collective layer of Hourglass model of the Grid. The discovery protocol is used for a searching of resources according to coarse grained requirements of end-user.

Since Grid Information protocol (GRIP) is used as both discovery and inquiry protocol, it is discussed below along with inquiry protocol.

### C. Resource Inquiry Protocol

In Grid environment, higher level services (users or schedulers) use inquiry protocol to obtain information about the Grid entities (e.g., resources or services) known to an Information Provider. Inquiry protocol is used for a direct lookup of information about the Grid entity (e.g., resource or service). The inquiry protocol is issued on the Information Provider. Moreover, the inquiry protocol is used to refine the set of resources from the discovered resources by considering exact or fine grained resource requirement criteria of end users.

*Example Protocol: The GRid Information Protocol (GRIP) of MDS-2 of GT-2.*

The GRid Information Protocol (GRIP) supports both discovery and inquiry. The discovery is supported via search capability and inquiry is supported via lookup capability of the GRIP protocol. Moreover, the specification in [41] places following requirements on the GRIP protocol. GRIP is the core of the resource layer function of Monitoring and Discovery Service (MDS). GRIP must supply a security model capable of supporting Grid Security Infrastructure−GSI (for security). GRIP must support a rich information model. GRIP must not require a specific storage model. GRIP must not require enumerable entities. GRIP must supply a query function. GRIP must be deployed universally. GRIP should support distributed operations. GRIP must not require a consistent global state.

To satisfy above stated requirements, GRIP was always implemented as LDAP [41]. The LDAP defines a data model for message structure and format; query language for search and inquiry operation; and wire protocol [38]. However, the LDAP query language cannot specify relational joins, and thus the join operation is not provided as part of basic GRIP query language. For this reason, some extension to LDAP should be adopted for join operations. Moreover, by placing a requirement "GRIP must not require a specific storage model", it becomes possible that the Information Provider is not required to store information about its entity(s) (e.g., resources or services). For example, the Information Provider can generate dynamic information (i.e., information can come directly from instruments, e.g., network bandwidth status, weather information, etc) only when it is requested.

## VIII. GRID MANAGEMENT PROTOCOLS

Grid management protocols are concerned with secure negotiation of resource use; brokering, scheduling, initiation, and control of job(s); and monitoring, control, accounting, and payment of used shared resources and jobs. In Grid, management is the process of monitoring an entity (resource), controlling it, maintaining it in its environment, and responding appropriately to any changes of internal or external conditions [42]. The resource management [43] [44] considers all the aspects for the remote job execution such as locating a capability, arranging for its use, utilizing it, and monitoring its state [18].

In Grid, resource management at global scale, called Grid resource management [44], relies on the local resource management [45]. The local resource management (LRM) [45] deals with the resource management of few resource types and the resources which reside within single administrative domain. Moreover, the local RMS maintains exclusive control over the resources. However, the underlying resources can have different local interfaces and protocols. For this reason, to enable management of such local resources at global scale, standard interface and protocol (such as GRAM) are provided for local resources by LRM. The Grid resource management (GRM) [44] deals with the management of heterogeneous resources present in different administrative domains at global scale. The GRM needs to provide transparent access to geographically distributed resources in highly dynamic environment with considering diverse resource management policies and different access models. Moreover, the GRM has no exclusive control over any resources, however for access of local resources, it rely on LRM and standards interfaces and protocols provided by LRM.

The characteristics of core Grid management protocols are summarized in Table V. The table shows at-the-glance look of these protocols, however, comprehensive discussion on the protocols related to Grid management activities is provided in the following sub-sections.

TABLE V CORE GRID MANAGEMENT PROTOCOLS, THEIR USAGE, AND THEIR CHARACTERISTICS

| Protocol | Example | Protocol Usage | Characteristics of Protocol |
|---|---|---|---|
| Grid Monitoring Protocol | GRid Information Protocol (GRIP) of MDS-2 of GT-2 | To provide dynamic information about the characteristics and status of resources of interest. | • Supports push and pull information delivery models of monitoring information.<br>• Represents monitoring information in LDAP format. |
| Job Submission and Job Management Protocol | Grid Resource Allocation and Management (GRAM) protocol of Globus | To provide secure and reliable creation and management of execution of remote computation-jobs. | • Uses GSI mechanism for authentication, authorization and credential delegation.<br>• GRAM of GT-2 uses Resource Specification Language (RSL) [45] for job description. While, GRAM of GT-4 uses XML based Job Description Document (JDD), expressed in JSDL [46], for job description.<br>• Provides standard interface to diverse local resource manager. |
| Advance Reservation and Co-allocation Protocol | Advance reservation protocol of Reality Grid | To reserve resources on a single Grid node in advance. | • Non-blocking protocol to avoid indefinite blocking of other consumers from booking the resource.<br>• Provides state management for reservation process.<br>• Defines protocol messages for interaction between consumer and provider.<br>• Provides agreed outcome at both consumer and provider at the termination of reservation process. |
|  | Co-allocation protocol of RealityGrid | To reserve multiple resources on different Grid nodes simultaneously for a single job. | • Provides cancellation of reservations at any time. No guarantee of atomicity.<br>• Specifies agreed outcomes between the user and the co-scheduler.<br>• Supports reservation in hierarchy. |
| SLA Negotiation Protocol | Service Negotiation and Acquisition Protocol (SNAP) | To provide coordinated resource management using generic resource management operations | • Defines Task Service-level Agreements (TSLAs), Resource Service-level Agreements (RSLAs), and Binding Service-level Agreements (BSLAs). |

| | |
|---|---|
| (submit, acquire, and bind). To provide negotiation of different SLAs. | • Supports different negotiation patterns to arrive at mutually agreed conclusion.<br>• SNAP defines how a service represents resource management functionality to the network |

### A. Grid Monitoring Protocol

Monitoring is the process of collecting information about the characteristics and status of resources of interest [47]. The Grid monitoring is about knowing how characteristic of resources varies over period of time rather than knowing characteristics of an entity at a specific point in time. Thus the monitoring can help in taking decision about application adaptation, performance diagnosis, fault detection, performance analysis, performance tuning, performance prediction, and scheduling.

We take GGF's Grid monitoring architecture and Globus toolkits' MDS for understanding of the monitoring. The GGF's Grid Monitoring Architecture [48] is scalable, efficient, and flexible architecture that supports interactions among three components: consumer, producer, and directory service. The consumer is any program that receives event data from producer. The producer provides event data to consumer either by request or asynchronously. The directory service contains listing of all available event data and their associated producers. Thus, the directory service allows information consumers (e.g., resource schedulers, users, programs, and visualization tools) to discover and understand the characteristics of information that is available. Moreover, no specific API, protocols or schema are defined by this architecture. However, LDAP for communicating with directory service and Simple Object Access Protocol (SOAP) for subscribing for event notification are used by many implementations.

*Example Protocol: GRid Information Protocol (GRIP) of MDS-2 of GT-2.*

MDS's GRIP is designed to support both discovery and monitoring. It supports both push and pull information delivery models [38] of information. In pull model, request-response exchange of protocol provides on-demand access to information. And, in push mode, an initial subscription request of protocol establishes subsequent asynchronous delivery. Thus, in GT-2 based Grid architecture, GRIP is used as monitoring protocol.

### B. Job Submission and Job Management Protocol

The job submission protocol deals with how job is represented in document form and how job is reliably transmitted to remote compute node. And, the job management protocol deals with how job is managed at remote Grid node, how the status of remote job is communicated to its user, and how user is allowed to control job remotely. Moreover, the user describes the job using job description language by specifying details such as the name of the executable, the working directory, where input and output should be stored, and the queue in which it should run. This description is sent to the remote resource through job submission protocol and ultimately it results in the creation of an instance job, which can be remotely managed using job management protocol. In short, the protocol includes job specification, message format, and sequence of message interactions between *user and job* and between *user and intermediary* for efficient management of job.

*Example Protocol: Grid Resource Allocation and Management (GRAM) protocol*

The GRAM is a component of Globus Toolkit [19], and its GRAM protocol provides facility of securely and reliably creating and managing execution of remote computation-jobs. Moreover, the GRAM is not a job scheduler, but rather a set of services and clients for communicating with a range of different local job schedulers and meta-schedulers using a common protocol called GRAM protocol. Consequently, the GRAM component itself does not implement any local resource management functionality but for this, it relies on interfaces provided by local resource management systems. The job in GRAM is managed by maintaining the state of job through state transition diagram.

In the GRAM protocol, GSI mechanisms and protocols are used for authentication, authorization, and credential delegation to remote computation-jobs. Specifically, the GRAM protocol in GT-2 uses (i) trusted "gate-keeper" process to initiate remote computation-jobs, (ii) a "job-manager" to manage the remote computation jobs, and (iii) a "GRAM reporter" to monitor the state of computation-job and publishing them to MDS-2. Specifically, the GRAM protocol, in GT-2, uses custom RSL syntax and custom parser, and custom https engine to submit and access the job, while, WS-GRAM, in GT-4, uses XML [49] based (Job Description Document) JDD expressed using Jobs Submission Description Language (JSDL) [46] and WSRF coherent protocols and hosting environment. (Interested readers are directed to [50] for getting more details on comparison between GT-2's GRAM and GT-4's GRAM).

### C. Advance Reservation and Co-allocation Protocol

Advance reservation is the process (initiated by resource consumer) of setting aside a resource (provided by resource provider) for defined time interval for use in future date [51] [52]. And, the co-allocation [53] is the process of booking multiple resources for a single job [52]. In this sub-section, we briefly discuss protocols related to these two activities.

There are two stages in advance reservation process, (i) query (ii) confirmation of reservation. In query stage, the resource consumer asks to resource provider to provide resources related information such as availability, associated cost, and reservation cancellation policy. Consequently, the resource provider provides this information to consumer. Consequently, if consumer agrees upon the information given by provider, then it confirms the reservation. To accomplish above stages and to control the reservation process according to ordering of messages exchanged by provider and consumer, there is a need to maintain status of advance reservation process in form of some finite state machine. Moreover, negotiation process can be used to alter properties of reserved resources. Moreover, advance reservation protocol must support the functionality of cancellation of reserved resources by any party (provider or consumer) at any time.

Co-allocation protocol is used to co-ordinate process of multiple advance reservations. The process is performed by Co-scheduler, which is an intermediary that reserves a set of resources on behalf of user. The user's request for required resources is given to the co-scheduler. And, co-scheduler co-ordinates the process of multiple advance reservations using a form of two phase commit protocol [54]. Thus, the co-allocation protocol can ensure that all the required resources by the user are reserved by resource managers, either all commit or all abort. Moreover, from resource point of view, it doesn't matter whether the reservation request has come from the consumer (user) or from the co-scheduler.

*Example Protocol: The advance reservation protocol* [52] *of RealityGrid* [55].

The advance reservation protocol models the advance reservation process as a single long running work. The advance reservation protocol is made non-blocking to avoid indefinite blocking of other consumers from booking the resource. For this reason, a quote sent by a provider has a limited life time, and if a currently negotiating consumer provides no decision within quote time period, any messages received afterwards may not be accepted by the provider. Moreover, the protocol assumes reliable messages delivery, i.e., messages are eventually delivered in the FIFO order to the destination and messages are delivered exactly once.

The protocol specifies five possible outcomes in which an advance reservation process may end. These outcomes are booking rejected, successful execution, faulted execution, consumer cancelled, and provider cancelled. Moreover, the protocol defines various messages (interested readers are directed to [52] for more details) that are exchanged between provider and consumer. It also defines the finite state machines for both consumer contract and provider contract in such a way that the consumer and provider services can reach the same agreed outcome at termination. Moreover, the consumer and provider contracts define the causal relationship (order) in which the consumer service and provider send and receive the protocol messages, respectively.

*Example Protocol: The Co-allocation protocol* [52] *of RealityGrid* [55].

The co-allocation protocol specifies the order in which the co-scheduler sends and receives messages with the user and each resource. The same messaging protocol, as defined in Advance Reservation Protocol, is used between the user and co-scheduler and between the co-scheduler and each resource.

The protocol specifies agreed outcomes between the user and the co-scheduler. These outcomes are successful and faulted execution, rejected reservation, user initiated cancellation, and co-scheduler initiated cancellation. The working of the co-allocation protocol is similar to two phase commit [54] protocol. However, the protocol can not support the atomicity property (all the required resources are reserved), as resource providers can cancel reservations at any time. During working of this protocol, the co-scheduler sends a booking request to each of the required resources. If all the required resources are successfully booked, then the user is informed of a successful booking. When the co-scheduler receives confirmation message (commit) from user, it sends a confirmation message to each of the resource to confirm the reservation. All the resources now become successfully reserved for user. Once the job completes at resource, it sends a close message to the co-scheduler, and specifies whether the job was successfully executed or faulted. Consequently, the co-scheduler informs the user of a successful execution if the job successfully executes at each resource. Otherwise, the co-scheduler informs the user that the job faulted.

There are cases in which reservation will not be successful. For instance, (i) The rejected reservation is possible if one or more of the required resources reject the reservation request, (ii) If reservation process is still active at co-scheduler, the user can cancel an advance reservation by sending cancel message to co-scheduler, (iii) The co-scheduler starts cancellation if one of the resource providers cancels the reservation. When a co-scheduler receives either a cancellation message, either from user or from resource providers, or rejection message, from resource provider, it cancels all the successful reservations done before on other resources.

### D. SLA Negotiation Protocol

Service Level Agreement (SLA) is an agreement between a service provider and service consumer that defines the terms under which a capability offered by a service can be used [56]. The terms generally include the Quality of Service (QoS) and the cost associated with the use of service. And, negotiation is a decision process in which two or more parties willingly make individual decisions and interact with each other for mutual gain [57]. The intention of negotiation process is to produce an agreement upon range of issues or courses of action, called as negotiation objects, of bargaining processes. This process continues till an agreement or deadlock is reached, or even one or more parties are not willing to continue in the process [58]. The negotiation protocol covers the permissible types of participants, negotiation states, the events which cause negotiation states to change and the valid actions of the participants in particular states [59]. Specifically, in SLA negotiation protocol, the SLAs work as negotiation objects and service provider and service consumer work as negotiating parties. The use of SLA negotiation protocol is to establish a mutual agreement between a resource provider and a resource consumer on QoS and cost associated with the use of service.

It was expected in [56] that the future Grid will need to support (i) Service Oriented Architecture, (ii) generalized resource management interfaces for different requirements such as task submission, workload management, on-demand access, co-allocation, co-scheduling, advance reservation, resource brokering , (iii) provisioned rather than best-effort service. The SLAs can perfectly address these requirements.

*Example Protocol: Service Negotiation and Acquisition Protocol (SNAP) [60]*

A generalized resource management framework, which is applicable to any type of resource, considers three basic resource management operations: submit, acquire, and bind, whose incremental composition can allow coordinated resource management for above stated three requirements. The SNAP protocol defines three kinds of SLA for three management operations. These SLAs are Task Service-level Agreements (TSLAs), Resource Service-level Agreements (RSLAs), and Binding Service-level Agreements (BSLAs).

- *TSLA [60]*: It is an SLA associated with performing a specified activity or task. TSLA characterizes a task in terms of its service steps and resource requirements.
- *RSLA [60]*: It is an SLA associated with acquiring a resource. The RSLA characterizes a resource in terms of its abstract service capabilities. In fact, the RSLA can be negotiated without specifying the activity or task for which the resource will be used.
- *BSLA [60]*: It is used to negotiate for using a resource for a task. As a result, an agreement to provide capability is bound to an agreement to perform a task using SLA.

The SNAP protocol should support different negotiation patterns such as propose/accept, (propose/counter propose)*/accept, and …/accept/commit to arrive at mutually agreed conclusion. During working of SNAP protocol, the service provider publishes or advertises its SLAs and types of negotiation patterns it is willing to support. And, the service consumer performs discovery of services supporting desired capability, then, the consumer uses these SLAs and negotiation pattern to negotiate with service provider to arrive at conclusion. Moreover, the SNAP defines how a service represents resource management functionality to the network. These SNAP protocol elements include: the time period for which SLA holds, the identity of the negotiating parties, the terms of agreement, etc.

## IX. Grid Interface Protocols

We, first, discuss basic concepts, terms, and terminology related to Web services. Then, we discuss important protocols that are used to provide open-standard based interoperable Grid interfaces.

- *Web service*: A Web service is a software system designed to support interoperable machine-to-machine interaction over a network [61]. To make interaction machine-processable, it uses (eXtensible Markup Language) XML [49] to tag and serialize the data; Simple Object Access Protocol (SOAP) to encapsulate the interaction messages; and Web Services Description Language (WSDL) to describe the services provided by it.
- *Service Oriented Architecture (SOA)*: SOA [62] is an architectural style, design style and a design principle for application development and integration development with the focus on using common, standard based, and open protocols to offer features such as modularity, encapsulation, loose coupling, separation of concerns, and composability to services to make them reusable without any restriction of platform, programming language. etc. Although the SOA concepts and principles have existed longer than the Web services, the Web services technology has popularized the realization of SOA.
- *Open Grid Services Infrastructure (OGSI)*: It is a specification by GGF's OGSI Working Group. OGSI [63] introduces the idea of stateful Web services and its management. OGSI provides detailed specification on creating, naming, managing lifetime, monitoring, grouping, and exchanging information among Grid services. The OGSI 1.0 specification [63] defines certain specific PortTypes (i.e., interfaces) that should be implemented by any Web service to become a Grid service
- *Open Grid Services Architecture (OGSA)*: It is a specification by GGF's OGSA Working Group. The OGSA [64] is an SOA based Grid architecture defined by a set of standards being developed within the GGF. The goal of OGSA is to standardize practically all the services that are commonly found in a Grid system (e.g., job management services, resource management services, security services, scheduling services, etc.) by specifying a set of standard interfaces for these services.
- *Grid-service*: Grid services [65] are Web services that conform to a set of conventions and support certain interfaces to make Web-services stateful. Two standards (i) OGSI based Grid services and (ii) Web Services Resource Framework (WSRF) based Grid services are available to implement OGSA-compliant Grid services. However, the WSRF based services supersede OGSI based services.
- *World Wide Web Consortium (W3C)*: The World Wide Web Consortium (W3C) [66] develops interoperable technologies (specifications, guidelines, software, and tools) to lead the Web to its full potential [66]. It has produced standards in various areas related to web. The major areas include HTML and style-sheets processing; XML and XML processing; Web services; and semantic web.
- *The Organization for the Advancement of Structured Information Standards (OASIS)*: OASIS [67] is a not-for-profit consortium that drives the development, convergence, and adoption of open standards for the global information society [67]. It has produced standards in various areas related to information structure. The major areas include Web services, Web services discovery, e-business, security, documentation markup, information exchange and interoperability, and XML Processing.

Protocols under Grid interface category are open standards based and they are used to provide standard interfaces between service consumers and service providers in Service Oriented Architecture (SOA) based Grid architecture. These protocols include: SOAP, WS-Notification, WS-ResourceFramework, WS-Addressing, WS-Security, and WS-SecureConversation. The characteristics of the core Grid interface protocols are summarized in Table V. The table shows at-the-glance look of these protocols, however, comprehensive discussion on the protocols related to Grid management activities is provided in the following sub-sections.

TABLE VI Core Grid Interface Protocols, their Usage, and their Characteristics

| Protocol | Example | Protocol Usage | Characteristics of Protocol |
|---|---|---|---|
| Basic Protocols for Web Service based Interaction | SOAP | To provide standard based interoperable message exchange between Web service and Web service client. | • W3C standard<br>• XML based simple, extensible, and lightweight mechanism for exchanging structured and typed information.<br>• Transport independent (can work on HTTP, SMTP, FTP, BEEP, etc.) |
| Protocols for Stateful Web Services | Web Services Resource Framework (WSRF) | To implement stateful services and their management. | • Standard by OASIS<br>• State representation of Stateful Web service in XML format.<br>• Defines monitoring and destruction of stateful Web service.<br>• Supports immediate and time-based destruction of stateful Web services.<br>• Defines grouping of stateful Web services.<br>• Defines a common way of representing faults. |
| | Web Services Addressing (WS-Addressing) | To provide transport-neutral mechanisms to address Web services and messages. | • Standardized by W3C<br>• Uses End Point Reference (EPR) to address Web services and messages. EPR can include extra information such as resource identifier for stateful Web services.<br>• Can support different Message Exchange Patterns (MEPs) such as request-response, solicit-response, one way, subscribe-notify, etc. |
| | Web services Notification (WS-Notification) | To provide asynchronous notification of events that occurs during lifetime of Web services. | • Standardized by OASIS<br>• Supports direct and brokered delivery of notification messages.<br>• Defines a mechanism to organize and categorize items of interest for subscription (known as "topics"). |
| Protocols for Web Services Security | WS-Security | To provide message level security | • Standardized by OASIS<br>• It supports for multiple security token formats, multiple trust domains, multiple signature formats, and multiple encryption technologies.<br>• Supports various token profiles such as Username token profile, X.509 token profile, Security Assertion Markup Language (SAML) token profile, Kerberos token profile, and Rights Expression Language (REL) token profile.<br>• Uses XML Signature to guarantee message integrity.<br>• Uses XML Encryption to provide message confidentiality. |
| | WS-SecureConversation | To provide efficient and reliable secure conversation during session between communicating parties. | • Standardized by OASIS<br>• Defines extensions to provide security context establishment and sharing, and session key derivation.<br>• Defines a new security token called Security Context Token (SCT) for secure conversation.<br>• Defines ways of establishing a security context. |

### A. Basic Protocols for Web Service based Interaction.

Currently, widely used protocol to access and invoke a function on remote server is Simple Object Access Protocol (SOAP) [68] [69]. SOAP provides XML [49] based simple, extensible, and lightweight mechanism for exchanging structured and typed information between peers in a decentralized and distributed environment. SOAP consists of three parts: (i) The SOAP envelope,(ii) The SOAP encoding rules, and (iii) The SOAP Remote Procedure Call (RPC) representation. The SOAP envelope construct defines an overall framework for expressing what is in a message, who should deal with it, and whether it is optional or mandatory. The SOAP message consists of a mandatory SOAP envelope, an optional SOAP header, and a mandatory SOAP body. The header is a generic mechanism for adding features to a SOAP message in a decentralized manner. The SOAP header can be processed by SOAP intermediaries, while SOAP body is processed by ultimate receiver of the message. The encoding rules define a serialization mechanism that can be used to exchange instances of application-defined data types and data structures. The SOAP RPC representation defines a convention that can be used to represent remote procedure calls and responses.

The service that is remotely accessible through Internet is called a Web service, as its invocation is done through Web based transport protocol such as Hyper Text Transport Protocol (HTTP) [70]. However, other protocols such as SMTP [71], FTP [24], BEEP [72], etc can also be used for transporting SOAP messages. Moreover, the capabilities of Web services are represented using a W3C standard based language called Web services Description Language (WSDL) [73]. The WSDL specification [73] defines WSDL as "*an XML grammar for describing network services as collections of communication endpoints capable of exchanging messages.*" Essentially, a WSDL document describes (i) how to invoke a service, (ii) structure of information that is being exchanged, (iii) the sequence of messages for an operation, (iv) protocol bindings, and (v) the location of the service.

*B. Protocols for Stateful Web Services*

Typically, Web services are implemented as stateless. However, in Grid computing the state of a resource or service is often important and may need to persist across interactions or transactions. While using a resource as a Web service, it becomes important to decide about how to create, destroy, name, and access Web services (or resources); how to provide state and lifetime management of Web services (or resources); how to manage group of Web services (or resources); and how to monitor state of Web service (or resource).

Before the convergence of the Web service community and the Grid community [74], OGSI 1.0 [63] was used to provide stateful Web services. OGSI 1.0 is based on WSDL 1.1, which does not support interface or PortType inheritance. OGSI 1.0 uses Grid WSDL (GWSDL) [63], which is a modified WSDL that allows interface extension and declaration of service data as part of Web service interface definition, to provide PortType inheritance. Hence, in this sense OGSI 1.0 deviated from WSDL. OGSI 1.0 treats a resource as a Web service itself by supporting the GridService PortType [63]. Because, the OGSI 1.0 had too much stuff in one specification and it was too much object oriented, it could not become popular among communities. This resulted into evolution [75] of OGSI 1.0 into three separate specifications.

Web Services Resources Framework (WSRF) [76] specification together with other related specifications: WS-Notification, and WS-Addressing, provides similar functionality to that of OGSI 1.0. WSRF specification supersedes OGSI and completes Grid and Web services convergence. These three separate specifications, as opposed to a single specification–OGSI 1.0, enable flexible composition of different functionality in an incremental or mix-match way. These three specifications can also be used independently.

*1) Web Services Resource Framework (WSRF)*

WSRF, an OASIS standard, makes explicit distinction between the "service" and the "resources" acted upon by that service by using the implied resource pattern of WS-Addressing specification. In implied resource pattern, the service and associated resources are identified by means of a single entity called an endpoint reference (EPR). Thus, using WS-Addressing, the requestor does not provide the stateful resource identifier as an explicit parameter in the body of the request message, as done in OGSI 1.0. However, using WS-Addressing, the endpoint reference (EPR) provides the means to point to both the Web service and the stateful resource as part of SOAP header. The framework defines how to declare, create access, monitor for change, and destroy the WS-Resource, which is composition of a Web service and a stateful resource, through conventional Web services mechanisms. WSRF is a set of five related specifications: (i) WS-Resource [77], (ii) WS-ResourceProperties (WS-RP) [78], (iii) WS-ResourceLifetime (WSRF-RL) [79], (iv) WS-ServiceGroup (WSRF-SG) [80], and (v) WS-BaseFaults (WSRF-BF) [81]. These protocol specifications are briefly discussed below.

- *WS-Resource [77]*: WS-Resource describes the relationship between a Web service and a resource in the WS-Resource Framework. It also defines the pattern by which resources are accessed through Web services, and the means by which WS-Resources are referenced.

- *WS-ResourceProperties [78]*: The state of a resource is modeled using resource properties. This specification standardizes the means by which the definition of the properties of a WS-Resource may be declared as part of the Web service interface. This specification also defines how to query or update the resource property values. Two ways: pull (query) and push (notification) are supported to access resource properties. The resource properties, which are included in resource properties document, describe the representation of a resource, however, the implementation of a resource can be in form of a file, or a database record, or in-memory data structure, etc.

- *WS-ResourceLifetime [79]*: The WS-ResourceLifetime specification standardizes the means by which a WS-Resource can be destroyed and the means by which the lifetime of a WS-Resource can be monitored. It defines two means of destroying a WS-Resource: immediate (or synchronous) destruction and time-based (or scheduled, for soft state management [39]) destruction. However, the specification does not prescribe the means by which a WS-Resource is created. The creation of resource (or WS-Resource) is handled using Factory design pattern [82].

- *WS-ServiceGroup [80]*: It defines how to represent and manage group of Web services or resources. A ServiceGroup is a heterogeneous by-reference collection of Web services. And this group is represented as resource. In this specification, the ServiceGroup membership rules, membership constraints and classifications are expressed using the resource property model.

- *WS-BaseFaults [81]*: It defines a common way of representing faults that occur during execution of Web service to have common understanding of fault messages. WS-BaseFaults defines an XML Schema [83] type for a base fault. It also defines how this base fault type is used within Web service interfaces.

*2) Web Services Addressing (WS-Addressing)[84]*

It provides transport-neutral mechanisms to address Web services and associated messages [84]. It defines extensible and reusable constructs for message addressing properties and endpoint references. The messages are directed to Web service using an Endpoint Reference (EPR). Apart from address of Web services, the EPRs can also contain extra information (e.g., in WSRF,

ReferenceProperties element to identify a resource; security; and renewable references information). The message addressing properties enable formation of different Message Exchange Patterns (MEPs) such as request-response, solicit-response, one-way, subscribe-notify, etc.

*3) Web services Notification (WS-Notification)*

In Grid environment, when resource properties change or when resource is destructed, a resource user is required to be informed asynchronously of this event, so that the user can take decision on further processing. For such asynchronous event based communication, Web services or other entities such as WS-Resource require standardized way of interaction. Notifications or events based communication among Web services using publish/subscribe messaging is standardized in WS-Notification standard. WS-Notification is a group of three standards: WS-BaseNotification [85], WS-BrokeredNotification [86], and WS-Topics [87]. WS-Notification supports direct or brokered delivery of notifications messages. Topics and Topic Spaces define a mechanism to advertise topics for subscription. Base Notification defines direct notification. Moreover, the WS-Notification defines and standardizes the role of brokers, publishers, subscribers, and consumers.

- *WS-BaseNotification [85]*: It defines the Web services interfaces, including standard message exchanges among them, for two basic roles in the notification pattern namely the NotificationProducer and NotificationConsumer [85] required in point-to-point notification.

- *WS-BrokeredNotification [86]*: To support advanced messaging features such as demand-based publishing, load-balancing, queuing of messages, aggregation of messages, distribution of messages, and filtering of messages, an intermediary between message publisher and subscriber called NotificationBroker is defined by WS-BrokeredNotification. This specification defines NotificationBroker interface and standard message exchanges.

- *WS-Topics [87]*: WS-Topics defines a mechanism to organize and categorize items of interest for subscription known as "topics" [87]. WS-Topics defines four topic expression (simple, concrete, full and X-Path) dialects that can be used as subscription expressions in subscribe request messages and other parts of the WS-Notification system. It also specifies XML model for describing metadata associated with topics [87].

## C. Protocols for Web Services Security

TLS/SSL provides point-to-point or transport layer security. However, to allow processing of secured SOAP messages (headers) by intermediaries, transport layer security does not work. Hence, message level security is required, and it is provided by WS-Security and WS-SecureConversation standards. WS-Security and WS-SecureConversation are discussed in following sub-sections.

*1) WS-Security*

WS-Security provides support for multiple security token formats, multiple trust domains, multiple signature formats, and multiple encryption technologies [88]. It provides three main mechanisms, not complete security solution, to solve the problem of authentication, integrity, and confidentiality for Web services. Moreover, it does not define how to do authentication, but rather it defines (in different profile documents) how to include different security tokens in SOAP messages. Specifically, at present, it specifies Username token profile [89], X.509 token profile [90], Security Assertion Markup Language [91] (SAML) token profile [92], Kerberos token profile [93], and Rights Expression Language (REL) token profile [94]. WS-Security also provides a standard set of SOAP extensions that provides message integrity and confidentiality. WS-Security uses XML Signature [95] to guarantee message integrity and XML Encryption [96] to provide message confidentiality. In short, WS-Security tells how to use other security technologies in Web services environment.

*2) WS-SecureConversation:*

While message authentication, as provided by WS-Security, is useful for simple or one-way messages, it may suffer from performance problem if used for multiple message exchanges during a session or it may suffer from several forms of attack [88] (e.g., message replay attack). These two problems are addressed by WS-SecureConversation. The WS-SecureConversation defines extensions to provide security context establishment and sharing, and session key derivation [97]. WS-SecureConversation works in conjunction with WS-Security, WS-Trust [98] and WS-Policy [99] to allow sharing of security contexts

Instead of including the same security credentials in each SOAP message, as done in WS-Security, WS-SecureConversation allows for an initial message exchange to establish a security context and then using it for subsequent message exchanges. WS-SecureConversation defines a new security token called Security Context Token (SCT), which can be used as a security token along with WS-Security. For a secure communication, a security context needs to be created and shared by the communicating parties before being used. The WS-SecureConversation defines three different ways of establishing a security context among the communicating parties: (i) Security context token created by a security token service, (ii) Security context token created by one of the communicating parties and propagated with a message, and (iii) Security context token created through negotiation/exchanges.

## X. CONCLUSIONS

We classified core Grid protocols according to activities handled by these protocols. The protocol categories are Grid network communication and Grid data transfer protocols, Grid information security protocols, Grid information protocols, Grid management protocols, and Grid interface protocols. In each protocol category, we considered core Grid protocols with emphasis on their distinguishing characteristics. We also characterized core Grid protocols along with taking applied protocol as an example wherever possible. We briefly described the placement of core Grid protocols in Hourglass model, which is a standard model of Grid architecture representation, of Grid architecture. We believe that the work presented in this paper will help not only to beginners but also to people who are working in research areas such as protocol architecture, service interfacing, and Grid in general. The reason behind including Grid protocols from earlier Grid generations is that the concepts used in earlier protocols are still valid in Service Oriented Architecture based Grid architecture. We briefly clarify this in next paragraph.

Data model used in earlier Grid systems was either proprietary or non-standard or platform specific. But, now data model uses standard, XML based, platform independent representation. The mechanisms used in earlier protocols such as soft state management, notification, asynchronous communication, atomic activity, service and resource discovery, service and resource registration, etc are now being included into Web service standards (WS-*). APIs to access Grid middleware services are now becoming on Web service based interfaces.

We believe that Web services based architecture of middleware will break the barrier of connecting different Grid middlewares through Internet. Thus making feasible, the Global Grid Computing across diverse organizations connected through Internet, which will result in Global Virtual Organization providing different commodity and provisioned based services. Grid protocols have evolved starting from the first generation of the Grid. However, still work in providing solution for services and protocols to support provisioned based services, with considering economical gain to both service provider and service user, attracts research community. Grid management activity demands great research work to consider issues of interoperable scheduling, brokering, service trading, service payment, service provisioning violations, etc.

## REFERENCES

[1] Open Grid Forum. [Online]. Available: http://www.ogf.org/
[2] I. Foster, "What is the Grid? - a three point checklist," *GRID Today*, vol. 1, no. 6, July 2002.
[3] Worldwide LHC Computing Grid. [Online]. Available: http://lcg.web.cern.ch/LCG/
[4] Enabling Grids for E-sciencE (EGEE). [Online]. Available: http://egee2.eu-egee.org
[5] European DataGrid project. [Online]. Available: http://www.edg.org/
[6] GARUDA. [Online]. Available: http://www.garudaindia.in/
[7] National e-Science Centre. [Online]. Available: http://www.nesc.ac.uk/
[8] Open Science Grid. [Online]. Available: http://www.opensciencegrid.org/
[9] distributed.net. [Online]. Available: http://www.distributed.net/
[10] D. P. Anderson, J. Cobb, E. Korpela, M. Lebofsky, and D. Werthimer, "Seti@home: an experiment in public-resource computing," *Commun. ACM*, vol. 45, no. 11, pp. 56–61, 2002.
[11] J. Rajkowski and K. Brayer, "The core protocol set for the global grid," *Military Communications Conference, 2001. MILCOM 2001. Communications for Network-Centric Operations: Creating the Information Force. IEEE*, vol. 1, pp. 512–518 vol.1, 2001.
[12] B. White, "Layered communications architecture for the global grid," *Military Communications Conference, 2001. MILCOM 2001. Communications for Network-Centric Operations: Creating the Information Force. IEEE*, vol. 1, pp. 506–511 vol.1, 2001.
[13] I. Foster and C. Kesselman, "Concepts and architecture," in *The Grid 2: Blueprint for a New Computing Infrastructure*, I. Foster and C. Kesselman, Eds. San Francisco, CA, USA: Morgan Kaufmann Publishers Inc., 2004.
[14] K. Nadiminti and R. Buyya, "Global Grid computing: Where are we today?" *Enterprise Open Source Journal*, vol. 2, pp. 26–29, November/December 2006.
[15] I. Foster, C. Kesselman, and S. Tuecke, "The anatomy of the grid: Enabling scalable virtual organizations," *International Journal of Supercomputer Applications*, vol. 15, no. 3, pp. 200–222, 2001.
[16] M. Roehrig, W. Ziegler, and P. Wieder, "Grid scheduling dictionary of terms and keywords," Global Grid Forum, http://www.ggf.org/documents/GFD.11.pdf, Grid Scheduling Dictionary WG GFD-I.11, November 2002.
[17] I. Foster, C. Kesselman, and S. Tuecke, "The open grid services architecture," in *The Grid 2: Blueprint for a New Computing Infrastructure*, I. Foster and C. Kesselman, Eds. San Francisco, CA, USA: Morgan Kaufmann Publishers Inc., 2004.
[18] K. Czajkowski, I. Foster, and C. Kesselman, "Resource and service management," in *The Grid 2: Blueprint for a New Computing Infrastructure*, I. Foster and C. Kesselman, Eds. San Francisco, CA, USA: Morgan Kaufmann Publishers Inc., 2004.
[19] Globus Toolkit. [Online]. Available: http://www.globus.org/
[20] J. Postel, "Transmission Control Protocol," RFC 793 (Standard), Sep. 1981, updated by RFCs 1122, 3168. [Online]. Available: http://www.ietf.org/rfc/rfc793.txt
[21] ——, "User Datagram Protocol," RFC 768 (Standard), Aug. 1980. [Online]. Available: http://www.ietf.org/rfc/rfc768.txt
[22] E. He, P. V.-B. Primet, and M. Welzl, "A survey of transport protocols other than standard TCP," Global Grid Forum, http://www.ggf.org/documents/GFD.55.pdf, Data Transport Research Group GFD-I.055, November 2005.
[23] B. Allcock, L. Liming, and S. Tuecke, "GridFTP: A data transfer protocol for the grid," Grid Forum, http://www.gridforum.org/documents/GF520Drafts/gridftp_intro_gf5.pdf, Grid Forum Data Working Group on GridFTP.
[24] J. Postel and J. Reynolds, "File Transfer Protocol," RFC 959 (Standard), Oct. 1985, updated by RFCs 2228, 2640, 2773, 3659. [Online]. Available: http://www.ietf.org/rfc/rfc959.txt
[25] "GridFTP: Protocol extensions to FTP for the grid," Global Grid Forum, http://www.ggf.org/documents/GFD.20.pdf, Tech. Rep., April 2003.
[26] I. Mandrichenko, W. Allcock, and T.Perelmutov, "GridFTP v2 protocol description," Global Grid Forum, http://www.ggf.org/documents/GFD.47.pdf, GridFTP WG GFD-R-P.047, May 2005.
[27] M. Horowitz and S. Lunt, "FTP Security Extensions," RFC 2228 (Proposed Standard), Oct. 1997. [Online]. Available: http://www.ietf.org/rfc/rfc2228.txt
[28] I. Foster, C. Kesselman, G. Tsudik, and S. Tuecke, "A security architecture for computational grids," in *CCS '98: Proceedings of the 5th ACM conference on Computer and communications security*. New York, NY, USA: ACM, 1998, pp. 83–92.
[29] N. Nagaratnam, P. Janson, J. Dayka, A. Nadalin, F. Siebenlist, V. Welch, I. Foster, and S. Tuecke, "Security architecture for open grid services," Global Grid Forum, http://www.ggf.org/ogsa-sec-wg/, GGF OGSA Security Workgroup GWD-I (draft-ggf-ogsa-sec-arch-01), June 2003.

[30] D. Cooper, S. Santesson, S. Farrell, S. Boeyen, R. Housley, and W. Polk, "Internet X.509 Public Key Infrastructure Certificate and Certificate Revocation List (CRL) Profile," RFC 5280 (Proposed Standard), May 2008. [Online]. Available: http://www.ietf.org/rfc/rfc5280.txt

[31] S. Kille, M. Wahl, A. Grimstad, R. Huber, and S. Sataluri, "Using Domains in LDAP/X.500 Distinguished Names," RFC 2247 (Proposed Standard), Jan. 1998, updated by RFCs 4519, 4524. [Online]. Available: http://www.ietf.org/rfc/rfc2247.txt

[32] T. Dierks and E. Rescorla, "The Transport Layer Security (TLS) Protocol Version 1.2," RFC 5246 (Proposed Standard), Aug. 2008. [Online]. Available: http://www.ietf.org/rfc/rfc5246.txt

[33] C. Adams and S. Lloyd, *Understanding Public-Key Infrastructure: Concepts, Standards, and Deployment Considerations*. Macmillan Technical Publishing, 1999.

[34] S. Tuecke, V. Welch, D. Engert, L. Pearlman, and M. Thompson, "Internet X.509 Public Key Infrastructure (PKI) Proxy Certificate Profile," RFC 3820 (Proposed Standard), Jun. 2004. [Online]. Available: http://www.ietf.org/rfc/rfc3820.txt

[35] B. Jacob, M. Brown, K. Fukui, and N. Trivedi, *Introduction to Grid Computing, IBM Redbooks*. IBM, December 2005.

[36] V. Welch, I. Foster, C. Kesselman, O. Mulmo, L. Peralman, S. Tuecke, J. Gawor, S. Meder, and F. Siebenlist, "X.509 proxy certificates for dynamic delegation," in *Proceedings of the Third Annual PKI Workshop*, 2004.

[37] M. Ahsanta, J. Basneyb, and O. Mulmoa, "Grid delegation protocol," in *UK Workshop on Grid Security Experiences*, July 2004.

[38] K. Czajkowski, S. Fitzgerald, I. Foster, and C. Kesselman, "Grid information services for distributed resource sharing," in *HPDC '01: Proceedings of the 10th IEEE International Symposium on High Performance Distributed Computing*. Washington, DC, USA: IEEE Computer Society, 2001, pp. 181–194.

[39] S. Raman and S. McCanne, "A model, analysis, and protocol framework for soft state-based communication," in *SIGCOMM '99: Proceedings of the conference on Applications, technologies, architectures, and protocols for computer communication*. New York, NY, USA: ACM, 1999, pp. 15–25.

[40] W. Yeong, T. Howes, and S. Kille, "Lightweight Directory Access Protocol," RFC 1777 (Historic), Mar. 1995, obsoleted by RFC 3494. [Online]. Available: http://www.ietf.org/rfc/rfc1777.txt

[41] M. Helm, "Grid information services for resource sharing," Global Grid Forum, GIS Working Group GWD-GIS-018-00, July 2001.

[42] "Resource management in OGSA," Global Grid Forum, http://www.ggf.org/documents/GFD.45.pdf, Common Management Model (CMM) WG GFD-I.045, Mar. 2005.

[43] K. Krauter, R. Buyya, and M. Maheswaran, "A taxonomy and survey of grid resource management systems for distributed computing," *Softw. Pract. Exper.*, pp. 135–164, 2002.

[44] F. Magoulès, T.-M.-H. Nguyen, and L. Yu, Eds., *Grid Resource Management: Towards Virtual and Services Compliant Grid Computing*. Boca Raton, FL 33487-2742: CRC Press, Taylor & Francis Group, 2009.

[45] K. Czajkowski, I. T. Foster, N. T. Karonis, C. Kesselman, S. Martin, W. Smith, and S. Tuecke, "A resource management architecture for metacomputing systems," in *IPPS/SPDP '98: Proceedings of the Workshop on Job Scheduling Strategies for Parallel Processing*. London, UK: Springer-Verlag, 1998, pp. 62–82.

[46] A. Anjomshoaa, F. Brisard, M. Drescher, D. Fellows, A. Ly, S. McGough, D. Pulsipher, and A. Savva, "Job submission description language (jsdl) specification, version 1.0," Global Grid Forum, http://www.ggf.org/documents/GFD.136.pdf, JSDL-WG GFD-R.136, July 2008.

[47] S. Zanikolas and R. Sakellariou, "A taxonomy of grid monitoring systems," *Future Gener. Comput. Syst.*, vol. 21, no. 1, pp. 163–188, 2005.

[48] B. Tierney, R. Aydt, D. G. W. Smith, M. Swany, V. Taylor, and R. Wolski, "A grid monitoring architecture," Global Grid Forum, http://www.ggf.org/documents/GFD.7.pdf, GGF Performance Working Group GFD-I.7, January 2002.

[49] Extensible Markup Language (XML) 1.0 (Fifth Edition). [Online]. Available: http://www.w3.org/TR/REC-xml/

[50] M. Feller, I. Foster, and S. Martin., "GT4 GRAM: A Functionality and Performance Study," in *TeraGrid Conference 2007*, Madison, WI, 2007.

[51] J. MacLaren, V. Sander, and W. Ziegler, "Advanced reservations:state of the art. grid resource allocation agreement protocol working group sched-graap-2.0," Global Grid Forum, Grid Resource Allocation Agreement Protocol Working Group sched-graap-2.0, June 2002.

[52] D. Kuo and M. Mckeown, "Advance reservation and co-allocation protocol for grid computing," in *E-SCIENCE '05: Proceedings of the First International Conference on e-Science and Grid Computing*. Washington, DC, USA: IEEE Computer Society, 2005, pp. 164–171.

[53] K. Czajkowski, I. Foster, and C. Kesselman, "Resource co-allocation in computational grids," in *HPDC '99: Proceedings of the 8th IEEE International Symposium on High Performance Distributed Computing*. Washington, DC, USA: IEEE Computer Society, 1999, pp. 219–228.

[54] J. Gray and A. Reuter, *Transaction Processing: Concepts and Techniques*. San Francisco, CA, USA: Morgan Kaufmann Publishers Inc., 1992.

[55] RealityGrid. [Online]. Available: http://www.realitygrid.org/

[56] I. Foster and C. Kesselman, Eds., *The Grid 2: Blueprint for a New Computing Infrastructure*. San Francisco, CA, USA: Morgan Kaufmann Publishers Inc., 2005.

[57] L. Thompson, *Mind and heart of the negotiator, second edition, the*. Upper Saddle River, NJ, USA: Prentice Hall Press, 2000.

[58] P. C. K. Hung, H. Li, and J.-J. Jeng, "WS-Negotiation: An overview of research issues," in *HICSS '04: Proceedings of the Proceedings of the 37th Annual Hawaii International Conference on System Sciences (HICSS'04) - Track 1*. Washington, DC, USA: IEEE Computer Society, January 2004, pp. 10 pp.–.

[59] N. R. Jennings, S. Parsons, C. Sierra, and P. Faratin, "Automated negotiation," in *5th International Conference on the Practical Application of Intelligent Agents and Multiagent Systems (PAAM-2000)*, Manchester, UK, 2000, pp. 23–30.

[60] K. Czajkowski, I. T. Foster, C. Kesselman, V. Sander, and S. Tuecke, "SNAP: A protocol for negotiating service level agreements and coordinating resource management in distributed systems," in *JSSPP '02: Revised Papers from the 8th International Workshop on Job Scheduling Strategies for Parallel Processing*. London, UK: Springer-Verlag, 2002, pp. 153–183.

[61] Web Services Architecture. [Online]. Available: http://www.w3.org/TR/ws-arch/

[62] E. Newcomer and G. Lomow, *Understanding SOA with Web Services (Independent Technology Guides)*. Addison-Wesley Professional, 2004.

[63] "Open grid services infrastructue (OGSI) version 1.0," Global Grid Forum, http://www.ogf.org/documents/GFD.15.pdf, Open Grid Services Infrastructue WG GFD-R-P.15, Jun. 2003.

[64] D. Berry, A. Djaoui, A. Grimshaw, B. Horn, F. Maciel, F. Siebenlist, R. Subramaniam, J. Treadwell, and J. V. Reich, "The open grid services architecture, version 1.5," Global Grid Forum, http://www.ogf.org/documents/GFD.80.pdf, Open Grid Services Architecture WG GFD-I.080, Jul. 2006.

[65] I. Foster, C. Kesselman, J. M. Nick, and S. Tuecke, "Grid services for distributed system integration," *Computer*, vol. 35, no. 6, pp. 37–46, 2002.

[66] World Wide Web Consortium (W3C). [Online]. Available: http://www.w3.org/

[67] OASIS:Advancing open standards for the global information society. [Online]. Available: http://www.oasis-open.org/

[68] SOAP Version 1.2 Part 1: Messaging Framework (Second Edition). [Online]. Available: http://www.w3.org/TR/soap12-part1/

[69] SOAP Version 1.2 Part 2: Adjuncts (Second Edition). [Online]. Available: http://www.w3.org/TR/soap12-part2/

[70] R. Fielding, J. Gettys, J. Mogul, H. Frystyk, L. Masinter, P. Leach, and T. Berners-Lee, "Hypertext Transfer Protocol – HTTP/1.1," RFC 2616 (Draft Standard), Jun. 1999, updated by RFC 2817. [Online]. Available: http://www.ietf.org/rfc/rfc2616.txt

[71] J. Klensin, "Simple Mail Transfer Protocol," RFC 5321 (Draft Standard), Oct. 2008. [Online]. Available: http://www.ietf.org/rfc/rfc5321.txt

[72] M. Rose, "The Blocks Extensible Exchange Protocol Core," RFC 3080 (Proposed Standard), Mar. 2001. [Online]. Available: http://www.ietf.org/rfc/rfc3080.txt

[73] Web Services Description Language (WSDL) Version 2.0 Part 1: Core Language. [Online]. Available: http://www.w3.org/TR/wsdl20/

[74] J. Joseph, M. Ernest, and C. Fellenstein, "Evolution of grid computing architecture and grid adoption models," *IBM Syst. J.*, vol. 43, no. 4, pp. 624–645, 2004.

[75] K. Czajkowski, D. F. Ferguson, I. Foster, J. Frey, S. Graham, T. Maguire, D. Snelling, and S. Tuecke, "From open grid services infrastructure to ws-resource framework: Refactoring & evolution," March 2004. [Online]. Available: http://www-106.ibm.com/developerworks/library/ws-resource/ogsi_to_wsrf_1.0.pdf

[76] Web Services Resource Framework WSRF) – Primer v1.2. [Online]. Available: http://docs.oasis-open.org/wsrf/wsrf-primer-1.2-primer-cd-02.pdf

[77] Web Services Resource 1.2 (WS-Resource). [Online]. Available: http://docs.oasis-open.org/wsrf/wsrf-ws_resource-1.2-spec-os.pdf

[78] Web Services Resource Properties 1.2 (WS-ResourceProperties). [Online]. Available: http://docs.oasis-open.org/wsrf/wsrf-ws_resource_properties-1.2-spec-os.pdf

[79] Web Services Resource Lifetime 1.2 (WS-ResourceLifetime). [Online]. Available: http://docs.oasis-open.org/wsrf/wsrf-ws_resource_lifetime-1.2-spec-os.pdf

[80] Web Services Service Group 1.2 (WS-ServiceGroup). [Online]. Available: http://docs.oasis-open.org/wsrf/wsrf-ws_service_group-1.2-spec-os.pdf

[81] Web Services Base Faults 1.2 (WS-BaseFaults). [Online]. Available: http://docs.oasis-open.org/wsrf/wsrf-ws_base_faults-1.2-spec-os.pdf

[82] E. Gamma, R. Helm, R. Johnson, and J. Vlissides, *Design Patterns: Elements of Reusable Object-Oriented Software*. Addison-Wesley, 1995.

[83] XML Schema Part 0: Primer Second Edition. [Online]. Available: http://www.w3.org/TR/xmlschema-0/

[84] Web Services Addressing 1.0 - Core. [Online]. Available: http://www.w3.org/TR/ws-addr-core/

[85] Web Services Base Notification 1.3 (WS-BaseNotification). [Online]. Available: http://docs.oasis-open.org/wsn/wsn-ws_base_notification-1.3-spec-os.pdf

[86] Web Services Brokered Notification 1.3 (WS-BrokeredNotification). [Online]. Available: http://docs.oasis-open.org/wsn/wsn-ws_brokered_notification-1.3-spec-os.pdf

[87] Web Services Topics 1.3 (WS-Topics). [Online]. Available: http://docs.oasis-open.org/wsn/wsn-ws_topics-1.3-spec-os.pdf

[88] Web Services Security: SOAP Message Security 1.1 (WS-Security 2004). [Online]. Available: http://docs.oasis-open.org/wss/v1.1/wss-v1.1-spec-os-SOAPMessageSecurity.pdf

[89] Web Services Security UsernameToken Profile 1.1. [Online]. Available: http://docs.oasis-open.org/wss/v1.1/wss-v1.1-spec-os-UsernameTokenProfile.pdf

[90] Web Services Security X.509 Certificate Token Profile 1.1. [Online]. Available: http://docs.oasis-open.org/wss/v1.1/wss-v1.1-spec-os-x509TokenProfile.pdf

[91] SAML Specifications. [Online]. Available: http://saml.xml.org/saml-specifications

[92] Web Services Security: SAML Token Profile 1.1. [Online]. Available: http://docs.oasis-open.org/wss/v1.1/wss-v1.1-spec-os-SAMLTokenProfile.pdf

[93] Web Services Security Kerberos Token Profile 1.1. [Online]. Available: http://docs.oasis-open.org/wss/v1.1/wss-v1.1-spec-os-KerberosTokenProfile.pdf

[94] Web Services Security Rights Expression Language (REL) Token Profile 1.1. [Online]. Available: http://docs.oasis-open.org/wss/v1.1/oasis-wss-rel-token-profile-1.1.pdf

[95] XML Signature Syntax and Processing (Second Edition). [Online]. Available: http://www.w3.org/TR/xmldsig-core/

[96] XML Encryption Syntax and Processing. [Online]. Available: http://www.w3.org/TR/xmlenc-core/

[97] WS-SecureConversation 1.4. [Online]. Available: http://docs.oasis-open.org/ws-sx/ws-secureconversation/v1.4/os/ws-secureconversation-1.4-spec-os.pdf

[98] WS-Trust 1.3. [Online]. Available: http://docs.oasis-open.org/ws-sx/ws-trust/v1.3/ws-trust.pdf

[99] Web Services Policy 1.5 - Framework. [Online]. Available: http://www.w3.org/TR/ws-policy